\documentclass[conference]{IEEEtran}

\usepackage{cite}
\usepackage{amsmath,amssymb,amsfonts}
\usepackage{algorithmic}
\usepackage{graphicx}
\usepackage{graphics}
\usepackage{textcomp}
\usepackage[usenames,dvipsnames,table]{xcolor}
\def\BibTeX{{\rm B\kern-.05em{\sc i\kern-.025em b}\kern-.08em
    T\kern-.1667em\lower.7ex\hbox{E}\kern-.125emX}}
\usepackage[hyphens]{url}

\usepackage[bookmarks=true,breaklinks=true,letterpaper=true,colorlinks,citecolor=blue,linkcolor=blue,urlcolor=blue]{hyperref}
\usepackage[normalem]{ulem}
\usepackage{xspace}
\usepackage{subcaption}
\usepackage{booktabs}
\usepackage{multirow}
\usepackage{pifont}
\usepackage{arydshln}

\newcommand{\thesystem}{Beehive\xspace}
\newcommand{\PANICUdpEcho}{CALM\xspace}
\newcommand{\enso}{Ens\={o}\xspace}

\newcommand{\VRthruputComparison}{1.14$\times$\xspace}
\newcommand{\VRlatencyComparison}{1.13$\times$\xspace}
\newcommand{\VRenergyMin}{2.07$\times$\xspace}
\newcommand{\VRenergyMax}{2.32$\times$\xspace}

\newif\ifshowcomment
\showcommentfalse

\ifshowcomment
\newcommand{\todoo}[1]{\textcolor{red}{[{TODO: #1}]}}
\newcommand{\pratyush}[1]{\textcolor{orange}{P: #1}}
\newcommand{\theano}[1]{\textcolor{blue}{TS: #1}}
\newcommand{\katie}[1]{\textcolor{Orchid}{K: #1}}
\else
\newcommand{\todoo}[1]{}
\newcommand{\pratyush}[1]{}
\newcommand{\theano}[1]{}
\newcommand{\katie}[1]{}
\fi

\makeatletter
\newcommand{\linebreakand}{%
  \end{@IEEEauthorhalign}
  \hfill\mbox{}\par
  \mbox{}\hfill\begin{@IEEEauthorhalign}
}
\makeatother

\begin{document}

\title{
\thesystem: A Flexible Network Stack for Direct-Attached Accelerators
}

\author{
  \IEEEauthorblockN{Katie Lim\IEEEauthorrefmark{1}, Matthew Giordano\IEEEauthorrefmark{1}, Theano Stavrinos\IEEEauthorrefmark{1}, \\
  Irene Zhang\IEEEauthorrefmark{2}, Jacob Nelson\IEEEauthorrefmark{2}, Baris Kasikci\IEEEauthorrefmark{1}, Thomas Anderson\IEEEauthorrefmark{1}}
  \IEEEauthorblockA{
    \{katielim, mgiordan, thst\}@cs.washington.edu, \\
    \{irene.zhang, jacob.nelson\}@microsoft.com, \{baris, tom\}@cs.washington.edu
  }
  \IEEEauthorblockA{
  \IEEEauthorrefmark{1}University of Washington, Seattle, WA, U.S.A.,
  \IEEEauthorrefmark{2}Microsoft Research, Redmond, WA, U.S.A.
  }
}


\maketitle

\begin{abstract}

%

Direct-attached accelerators, where application accelerators are directly
connected to the datacenter network via a hardware network stack, 
offer substantial benefits in terms of reduced latency, CPU overhead, and energy use.
However, a key challenge is that modern datacenter network stacks are complex, with interleaved
protocol layers, network management functions, and virtualization support. 
To operators, network feature agility, diagnostics, and manageability are often considered
just as important as raw performance.
By contrast, existing hardware network stacks only support basic protocols and are 
often difficult to extend since they use fixed processing pipelines.


We propose \thesystem, a new, open-source FPGA network stack for
direct-attached accelerators designed to enable flexible and adaptive
construction of complex network functionality in hardware. Application and network 
protocol elements are modularized as tiles over a network-on-chip
substrate. Elements can be added or scaled up/down to match workload
characteristics with minimal effort or changes to other elements.  
Flexible diagnostics and control are integral, with tooling to ensure deadlock safety.  
Our implementation interoperates with standard Linux TCP and UDP
clients, with a 4x improvement in end-to-end RPC 
tail latency for
Linux UDP clients versus a CPU-attached accelerator. Beehive is available at \url{https://github.com/beehive-fpga/beehive}

\end{abstract}

\section{Introduction}
\begin{figure}[t]
	\includegraphics[width=\linewidth]{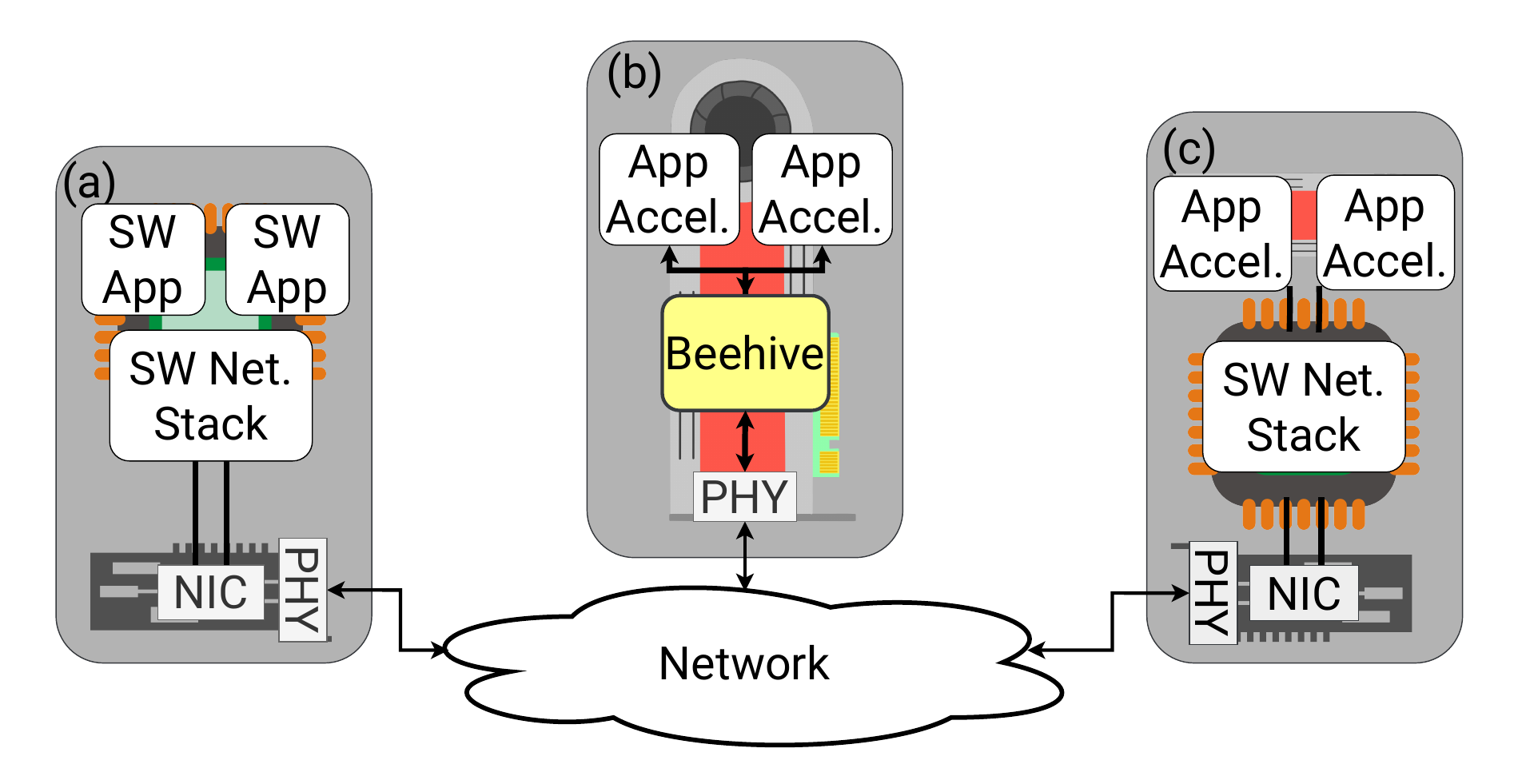}
	\caption{\textbf{(a)} represents a standard CPU server node; 
	\textbf{(b)} a direct-attached accelerator using the 
	\thesystem{} network stack; \textbf{(c)} an accelerator using a CPU
	network stack. 
	}
	\label{fig:setting_overview}
	\vspace{-0.6cm}
\end{figure}

Hardware accelerators are becoming increasingly common in datacenters to reduce
cost, improve performance, and reduce energy consumption relative to server
CPUs.  Typically, accelerators are hosted over the PCIe I/O bus, with the server
CPU mediating all communication with the accelerator, illustrated
in Figure \ref{fig:setting_overview}(c). 
An alternative model directly attaches the accelerator to the
network, with its own network functionality implemented in hardware, illustrated
in Figure \ref{fig:setting_overview}(b). 
Bypassing the CPU potentially reduces end-to-end latency, latency variability, and overhead, 
freeing up the CPU for other purposes.  

A barrier to any hardware network implementation is the difficulty of meeting the full set of 
datacenter network operational requirements~\cite{google_snap,azurerdma}.  
Network manageability, diagnostic visibility, and interoperability are often non-negotiable 
requirements, made more complex by the rapid evolution in host network stacks to meet application
and operational needs.
Beyond core protocols, such as TCP/IP, modern applications require higher-level 
functionality like remote
procedure call (RPC) processing, quality-of-service (QoS)
management~\cite{aequitas,breakwater}, encryption~\cite{ktls,mtls}, 
application-specific load balancing~\cite{istio_l7_lb,cilium_l7_lb}, 
and information flow control~\cite{gdpr}. 
Deployment flexibility necessitates management features like virtual
networking~\cite{openvswitch_network_virt_nsdi,microsoft_vfp_nsdi,google_andromeda_nsdi},
access control lists~\cite{hyperv_acl}, congestion control~\cite{swift,hpcc},
traffic prioritization~\cite{homa,bfc}, and load
balancing~\cite{ananta_sigcomm,maglev_nsdi,facebook_katran,cloudflare_unimog}.
Deployment maintainability requires dynamic support for network
monitoring~\cite{pigasus_osdi,ballard2010extensible},
reconfiguration~\cite{rmt_sigcomm,kim2013improving}, and
debugging~\cite{tammana2016simplifying}.  

An example of a highly-flexible software network stack is Google's Snap networking system~\cite{google_snap}. 
It is designed around composable message-passing engines,
with modules for load balancing, network virtualization, network management,
and custom transport protocols. 
New modules can be easily inserted anywhere 
in the stack, without re-engineering the rest of the stack.
Our question is whether we can do something similar in hardware.
Existing hardware network stacks are typically designed to support only a single 
application with minimal protocol complexity. 
Although some recent work has focused on flexible packet-level processing 
in hardware~\cite{clicknp_sigcomm,panic_osdi}, 
our aim is to support
flexibility across the entire network stack, including 
transport and application protocols. Other work has looked at hardware
offload of transport protocols, but these systems lack a range of essential network 
functions~\cite{limago,chelsio_t6,tonic}, or in the case of RDMA, require  
extensive engineering to make work in practice~\cite{azurerdma,redmark}.

This paper explores the design of an FPGA network stack that can realize
the benefits of direct-attached accelerators while supporting the extensibility,
incremental scalability, and manageability needed for production use.  
Flexibility is needed at multiple points in the network stack: in packet processing (layer 3), 
transport and congestion control (layer 4), the application layer (layer 7),
and in control/diagnostics operating alongside, and using, the data plane. 
Adding new functionality, differentially scaling protocol elements to meet application throughput needs, 
or inserting a new load balancing policy should be simple, as it is in software, without the need to disrupt or
re-engineer other layers. 


We propose and implement \thesystem, an open-source hardware network
stack architected as a collection of protocol functions that
communicate via message-passing over a scalable network-on-chip (NoC). We
provide automated tooling for managing differential scaling and load balancing of 
protocol elements, a control plane for diagnostics monitoring,
and compile-time deadlock analysis.  To make our design concrete, we implement
Ethernet, IP, UDP, TCP, network address translation (NAT), IP-in-IP
encapsulation, and additional support for control and debugging of network
functions. Our implementation interoperates with Linux TCP and UDP clients,
allowing unmodified remote procedure call (RPC) clients to use our accelerator. 

For our evaluation, we implement \thesystem and evaluate it on FPGA. We show that it offers a
4$\times$/1.5$\times$ improvement in end-to-end client RPC tail latency over Linux/user-level
TCP relative to mediating accelerator traffic through the server CPU, and up to
31$\times$ higher per-core throughput than a state-of-the-art CPU kernel-bypass
stack on small messages.

We implement two example applications using \thesystem: erasure coding as a bandwidth-oriented application and distributed consensus as a latency-sensitive application. 
First, modern datacenter storage systems often use erasure coding for better storage efficiency than
replication with comparable fault tolerance. 
We implement an erasure coding accelerator in \thesystem and show that, compared to a CPU-only version, the accelerator scales out to 62 Gbps using~20$\times$ less energy. 
Second, we show that accelerating a key piece of distributed consensus in hardware can reduce end-to-end median operation latency by \VRlatencyComparison, with \VRthruputComparison~ better per-core throughput and ~2$\times$ less energy than the CPU-only version.  

\begin{figure}[t]
    \includegraphics[width=\linewidth]{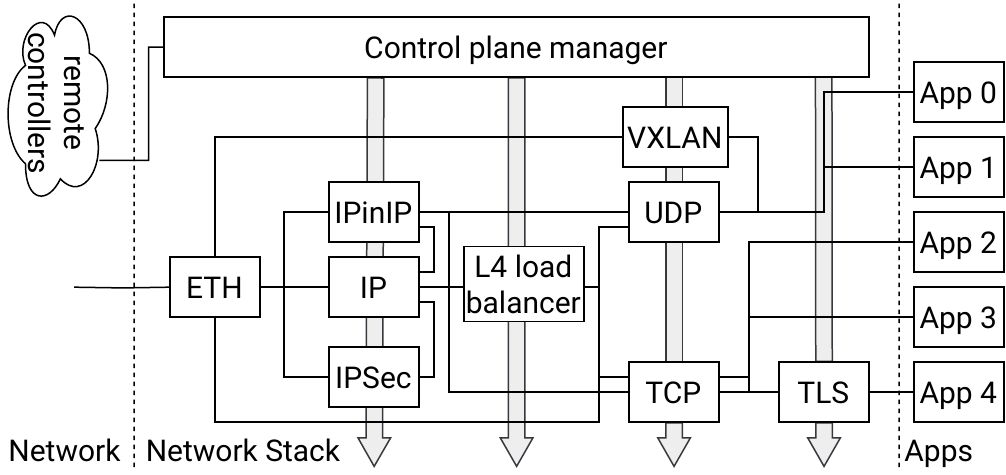}
    \caption{A high-level diagram of the type of network stack \thesystem{}
    targets. Along with multiple transport protocols, this stack has IP-in-IP
    and VXLAN for network virtualization and a component for an L4 load
    balancer. The downward arrows represent control-plane communication, which
    potentially needs access any module internal to the network stack.}
    \label{fig:high_level_net_stack}
\end{figure}

\noindent In summary, we contribute:
\begin{itemize}
	\item \thesystem, a design framework to build efficient and complex hardware
	      network stacks for direct-attached accelerator deployments in modern
	      datacenters.
	\item An open-source FPGA implementation of \thesystem that includes tools
	      and reusable components to build network stacks for 
	      accelerators that use different transport protocols, network
	      virtualization, and layer 7 functionality.
	\item A demonstration of \thesystem's ability to support scalability,
		  flexibility, low latency, high throughput, and energy efficiency by
		  integrating and evaluating an erasure coding accelerator and a consensus
		  accelerator.
\end{itemize}

\section{Motivation}
\label{sec:network-stack-placement}
We motivate direct-attached accelerators by investigating their latency benefits
over CPU-attached accelerators. Prior work has shown benefits over the Linux
network stack~\cite{catapult_v2, ibm_cloudfpga_measured}. However, cutting-edge
systems aiming for the lowest possible latency typically use a DPDK network
stack, which can achieve single digit microsecond
latencies~\cite{demikernel_sosp,erpc,fstack}.

Our experiment compares the direct-attached configuration in
Figure~\ref{fig:setting_overview}(b) and the software-hosted configuration in
Figure~\ref{fig:setting_overview}(c). We evaluate the performance of UDP echo,
where the client sends a UDP packet to a server and waits for the response
packet before sending another. We use Linux and F-Stack~\cite{fstack}, a DPDK
network stack, as software network stacks. We run 1,000,000 requests and measure
the round-trip time (RTT) for each request. 

For the direct-attached configuration, we use \thesystem implementing a UDP echo
server. We try both Linux and F-Stack as the clients. For the software-hosted
configuration, we use either the Linux network stack or F-Stack as the software
network stack and \enso~\cite{enso_osdi23} as the FPGA accelerator. \enso is an
FPGA-based NIC designed for efficient NIC-CPU communication over PCIe.
Internally, we tie \enso's network output to its input, so it operates as a loopback. For
software-hosted configurations, the client and server machines run the same
software stack (Linux or F-Stack).

\def\thickhline{\noalign{\hrule height.10pt}}

\begin{table}[t]
    \centering
    \caption{Comparison of median and p99 round-trip time of a UDP echo
    across different configurations. Client machines use software networking.
    \thesystem represents the configuration in
    Figure~\ref{fig:setting_overview}(b); Linux and DPDK to Accel. represent
    Figure~\ref{fig:setting_overview}(c).} 
    \resizebox{\linewidth}{!}{
        \begin{tabular}{l|cc|cc}
        \toprule
        Client & \multicolumn{2}{c|}{Linux Client}           & \multicolumn{2}{c}{DPDK Client}           \\ 
        \cline{1-5}\noalign{\vspace{0.4ex}}
        Server & \thesystem & \begin{tabular}[c]{@{}c@{}}Linux\\ to Accel.\end{tabular} & \thesystem & \begin{tabular}[c]{@{}c@{}}DPDK \\ to Accel.\end{tabular} \\ 
        \midrule
        Median Latency ($\mu$s) & 11.6  & 17.6 & 4.08 & 6.22 \\
        p99 Latency ($\mu$s)    & 15.3  & 61.2 & 4.43 & 6.79  \\
        \bottomrule
        \end{tabular}
    }
    \label{tab:netstack-experiment}
\end{table}

We report median and 99$^{th}$ percentile (p99) RTTs in
Table~\ref{tab:netstack-experiment}. As expected, trampolining every RPC through
the CPU on the way to the FPGA is both slower, and more variable, than when
the FPGA is directly attached to the network using \thesystem.  
When the network stack is provided by Linux, message latency can be affected
by CPU scheduling contention, so that \thesystem has 4$\times$ better p99 tail latency
than redirection through the CPU on this benchmark, and 1.5$\times$ better median latency.
When the network stack is at user level on both the client and server,
scheduling variance is reduced as the server CPU busy-waits for incoming requests,
at the cost of higher CPU overhead.  However, the relative benefit of \thesystem
is similar, with 1.5$\times$ better median and p99 tail latency relative to redirection
through the CPU.


This shows that even with a DPDK stack, direct-attached accelerators
can still provide a latency improvement, and the relative improvement is larger for 
tail latency compared to the Linux network stack. 
With this in mind, direct-attached accelerators are an appealing option. 
Realizing this benefit requires a hardware network stack that can be
flexibly reconfigured to meet the needs of datacenter network management.



\section{Design Goals} 
Our overarching goal for \thesystem is to
build an open-source FPGA hardware design to support emerging applications for
direct network-attached accelerators in a production environment. Figure
\ref{fig:high_level_net_stack} shows a high-level diagram of the type of network
stack architecture we want to be able to support. Applications may only use some
subset of these protocols and network functions. We now discuss our specific design goals.


\subsection{\thesystem Goals}
\label{sec:goals}
\noindent\textbf{Standard client protocols.} The vast majority of distributed
applications that might benefit from the availability of hardware acceleration
are designed to communicate using standard protocols such as IP, TCP, and remote
procedure call (RPC). Our framework needs to be able to support unmodified
client application and client host software communicating with the accelerator
using these standard protocols.

\noindent\textbf{Modularity.} However, network stacks are not fixed.
Requirements are constantly changing with new custom protocols (e.g.  Google's
Pony Express \cite{google_snap} or 1RMA \cite{1rma_sigcomm}) and network
functions. In order to facilitate rapid development and customization of the
network stack, our framework must be modular, so we can compose or integrate new
components with minimal to no modifications to existing components.




\noindent\textbf{Scalability.}
Building a complex network stack potentially means supporting a variety of
different components in the same design. Different components may be a bottleneck
depending on the application workload.  Thus, the architecture should be able to 
duplicate and scale out individual components, whether application or protocol logic,
as needed.


\noindent\textbf{Performance overhead and predictability.} Since performance and
performance predictability are key motivations to offload the network stack, the
stack should be able to deliver end-to-end application bandwidth at 100 Gbps
with minimal jitter if the accelerators have the capacity to support it.

\noindent\textbf{Management flexibility.} Components in a network stack need to
be able to interact beyond just passing packet data. For example, components
need to be able to expose interfaces to the control plane for telemetry and
debugging~\cite{envoy}. The control plane may also need to update state used by
a protocol or network function, such as configuring the load balancer used to
parcel work across application accelerator instances. Such configurability
should be possible even in large designs without extensive manual optimization.

\vspace{-.1cm}
\subsection{Comparing versus related work}
\vspace{-.05cm}
As shown in Table~\ref{tab:rel_work_table}, other related work does not meet all these goals. 
In terms of complexity, the Limago, a TCP engine written in Vitis HLS, is the closest to \thesystem.
However, it is not designed to allow for addition or replication of components within the stack, so it is limited in scalability and modularity. We discuss FPGA utilization comparisons further in Section \ref{sec:hw_util}. Unfortunately, we were unable to run Limago on FPGA using their code~\cite{limago_build_repo} to evaluate its performance, because the QSFPs did not come up on the FPGA board.

PANIC and ClickNP are the most similar architecturally to \thesystem as they are both based on message-passing over an interconnect, leading to similar performance and modularity benefits as \thesystem. 
Their implementations do not provide standard protocol support directly, but they could be extended to support the logic needed for these protocols.
Additionally, their interconnects can limit their scalability. While working on the experiment in Section \ref{sec:beehive_udp_microbench}, we found PANIC's crossbar was unable to support more than 8 endpoints, 4 of which are always used by its infrastructure. In ClickNP, components are directly connected using FIFOs, potentially causing fan-out issues when duplicating components. Because ClickNP is not open-source, we were unable to compare to it directly. 

\newcommand{\yes}{\textcolor{green!80!black}{\ding{51}}}
\newcommand{\nah}{\textcolor{red}{\ding{55}}}
\newcommand{\meh}{\textcolor{black}{\ding{83}}}

\newcommand{\specialcell}[2][c]{%
\begin{tabular}[#1]{@{}c@{}}#2\end{tabular}}

\newcommand{\specialcellL}[2][c]{%
\begin{tabular}[#1]{@{}l@{}}#2\end{tabular}}

\def\thickhline{\noalign{\hrule height.7pt}}

\begin{table}[t]
    \centering

    \resizebox{\linewidth}{!}{
\begin{tabular}{lcccccc}
	 & \specialcell{Std.\\Protocols} & Modular & Scalable & Performant & \specialcell{Mgmt.\\Features} & \specialcell{Open\\Source} \\
        \midrule
Limago~\cite{limago}       & \yes                  & \meh          & \nah           & \yes          & \nah          & \yes           \\
PANIC~\cite{panic_osdi}                                 & \meh                  & \yes          & \nah           & \yes          & \nah          & \meh           \\
ClickNP~\cite{clicknp_sigcomm}                               & \meh                  & \yes          & \meh           & \yes          & \yes          & \nah           \\
LTL~\cite{catapult_v2}                                  & \nah                  & \nah          & \meh           & \yes          & \meh          & \nah           \\
\thesystem                            & \yes                  & \yes          & \yes           & \yes          & \yes          & \yes           \\
        \bottomrule
\end{tabular}
    }
	\caption{\thesystem and prior work versus the goals in Section~\ref{sec:goals}. The stars indicate partially meeting the goal.}
    \label{tab:rel_work_table}
\end{table}


\vspace{-.1cm}
\section{Design}
\vspace{-.05cm}

\begin{figure}[t]
    \centering
    \includegraphics[width=.4\linewidth]{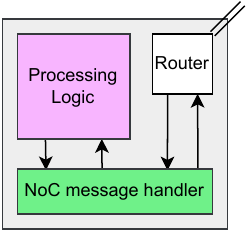}
    \caption{Architecture of a tile.}
    \label{fig:beehive_tile}
    \vspace{-0.5cm}
\end{figure}

\subsection{\thesystem's architecture}
\vspace{-.05cm}
The basic component in \thesystem{} is the tile, shown in
Figure~\ref{fig:beehive_tile}. Each tile has a network-on-chip (NoC) router,
some logic that handles NoC message construction and deconstruction, and some
processing logic, such as a protocol layer, network function, or application.
Tile routers are connected together to form the NoC topology. We do not require
a particular topology, although our prototype uses a 2D mesh. We require that the NoC
is reliable, point-to-point ordered, and uses deterministic, deadlock-free
routing.

A network packet is processed or constructed by passing NoC messages through a
chain of tiles. A NoC message consists of one header flit followed by some
number of body flits. The header flit typically contains data only relevant to
NoC-level routing, such as source and destination tile coordinates or number of
body flits. The body flits typically consist of both metadata flits containing
packet header fields and a number of data flits carrying unprocessed packet
payload. 

Each tile hop is responsible for determining the next tile that a message should
be sent to. This design is in contrast to earlier work which assumes that routes
can be fully determined on packet arrival~\cite{panic_osdi}.  We discuss this
decision in more detail in Section~\ref{sec:beehive_stream_routing}. This
component may vary in complexity from a static CAM to more complex logic, such
as content-based routing. The set of possible message chains is known ahead of
time for deadlock analysis, described in Section~\ref{sec:deadlock}.

\begin{figure}[t]
    \centering
    \includegraphics[width=\linewidth]{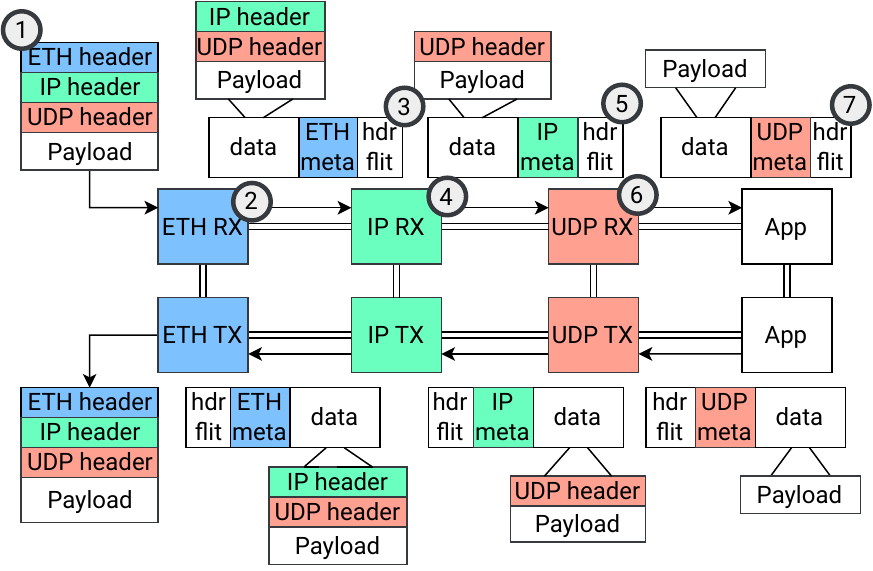}
    \caption{The flow through which a packet is processed or constructed in \thesystem{}.}
    \label{fig:beehive_basic_ex}
    \vspace{-0.7cm}
\end{figure}


\subsection{Processing a packet}
Figure~\ref{fig:beehive_basic_ex} shows an example of a basic UDP stack in
\thesystem, with a UDP packet moving through the receive and send paths. 
On the receive side, an Ethernet frame enters the Ethernet tile, which has ports
for the I/O from the transceivers in addition to the ports connecting to other
tiles. The processing logic within the tile parses and removes the Ethernet
header, realigning the data. This is then turned into a NoC message consisting
of a header flit, a metadata flit with the parsed Ethernet header, and some
number of data flits containing the remaining packet data.  The routing
component in the Ethernet tile uses the type field in the Ethernet header to
determine that the message should be passed to the IP tile. The IP tile
similarly parses the IP header, validates the header's checksum, and then
creates a NoC message to be sent to the UDP layer.
Finally, the UDP tile parses the UDP header, validates the packet's checksum,
and generates a NoC message to be sent to the application based on the port in
the UDP header.  The transmit path runs similarly, except instead of parsing
headers from the data flits, headers are added by each
protocol tile. After the Ethernet tile adds on the Ethernet header, it is sent
out the ports for I/O with the transceivers. This incremental composability is
good for our goal of modularity as it makes it easier to insert new
functionality between stages.

While there is only one possible destination for the tiles in this design, there
can potentially be multiple endpoints, such as other protocols (e.g. TCP
connected to IP), network services (e.g. network virtualization), or replicated
tiles for higher bandwidth. With replicated tiles, there are multiple ways to
decide on which tile should receive an incoming packet. The simplest method is
to distribute packets between them in a round-robin fashion. However, more
complex scheduling may be necessary if a tile holds state for particular flow.
In this case, it is important that packets from the same flow always go to the
same tile. This distribution can either be integrated within a tile or placed in
a dedicated tile. We discuss how we distribute packets to
duplicated tiles in Section \ref{sec:beehive_integration}.

\subsection{Message-passing interconnect} 
Being able to compose elements is essential for facilitating customization. 
We opt for a message passing model. This is beneficial for modularity, because
defining a message-passing format allows us to standardize the physical
interconnection between components, a recognized benefit in SoC design
\cite{route_packets_dac}, and makes it easier to chain offloads together. ClickNP
\cite{clicknp_sigcomm} and PANIC \cite{panic_osdi}, two modular packet
processing frameworks, have also used a message-based approach. The message
passing can be done over dedicated connections, which is the approach used by
ClickNP, or a NoC which is used by PANIC. 

We prefer a NoC interconnect for two main reasons related to our goal of scalability.
First, we can take advantage of the multiplexing provided by the NoC routers.
Certain tiles may interact with many other tiles, e.g. if we instantiate multiple copies of the same component or common services such as memory buffer storage.
Direct connections can lead to large multiplexers and wires with significant fan-out.
Although we could create specialized pipelined multiplexers and arbiters, these essentially look like NoC routers.

Second, we would like the interconnect wiring to remain stable whenever
possible. 
In the ClickNP model, top-level wires are determined by the
computational graph. 
If we wish to form a chain that links together two components that did not communicate before, we must add new interconnect wires, which are typically the longest wires. 
A NoC allows us to reuse physical wiring to chain any elements that exist in the design, as long as we are careful with deadlock.

These scalability benefits apply both to the data plane and control plane. We discuss the benefits further for the control plane specifically in Section~\ref{sec:beehive_ctrl_plane_design}.

\subsection{Tile chain routing}\label{sec:beehive_stream_routing}
In addition to NoC-level routing, \thesystem{} routes at the network packet
level to determine the sequence of tiles that need to be chained together. We
considered two routing methods: node-table routing, where each tile determines
the correct next tile, and source routing, where the chain of tiles is
completely determined when the first NoC message in the chain is created, such
as when a packet is first received from the network. We use node-table
routing, because certain classes of traffic we want to support for 
interoperability require per-flow state or non-trivial protocol processing to
fully determine the chain of tiles. 

Specifically, we consider routing for traffic that is either encrypted or is for layer 7. 
Encryption may obfuscate parts of packet payloads that are needed to fully route a packet, which would require the ingress tile to handle the decryption. 
An application request can span multiple packets.
Which application tile should receive an RPC may depend on the RPC header or even the contents of the request.
Further, the packets of one request, which may not fit in the first packet, could be reordered or interleaved with other requests.
To properly route such requests, an ingress tile would need to assemble or reorder the stream, further complicating the implementation.
In both cases, the ingress tile would need to implement significant, high-level protocol logic which is detrimental for modularity.

\begin{figure}[t!]
    \centering
    \begin{subfigure}[t]{\linewidth}
        \includegraphics[width=\linewidth]{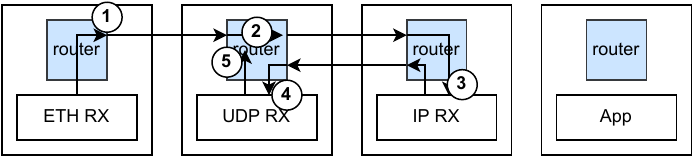}
        \caption{This tile assignment deadlocks due to the order that the packet needs
        to be processed in versus the order of the NoC links it traverses.}
        \label{fig:simple_deadlock}
    \end{subfigure}
    \hfill{}
    \begin{subfigure}[t]{\linewidth}
        \includegraphics[width=\linewidth]{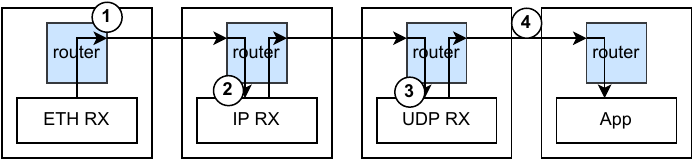}
        \caption{This tile assignment is able to avoid deadlock since the packet
        acquires NoC resources in order.}
        \label{fig:simple_nondeadlock}
    \end{subfigure}
    \caption{An example of how tile assignment affects deadlock. \thesystem
    takes advantage of protocol layer ordering, so a packet always acquires NoC
    resources in the same order.}
    \label{fig:deadlock_exs}
    \vspace{-.6cm}
\end{figure}

\subsection{Deadlock}\label{sec:deadlock} As with any NoC-based design, avoiding
message-based deadlock must be a consideration. We note that NoC deadlock
detection, avoidance, and recovery is a complex problem with a whole body of
research behind
it~\cite{dally_deadlock_free_routing,message_deadlock_recovery_avoidance_lankes_nocs,custom_message_deadlock_free_vlsi,anjan1995efficient,seiculescu2010method}.

NoCs can deadlock in two ways: at the routing level and at the message passing
level. To prevent routing-level deadlocks, we employ dimension-ordered
routing~\cite{dally_deadlock_free_routing}. Message passing deadlocks are a
bigger concern in \thesystem, because any tile can route to any other
tile at runtime. This means that our routing resources can get exhausted. The
deadlock in Figure~\ref{fig:simple_deadlock} is an example of this, in which the
UDP RX tile must route east twice in one chain, and it cannot route east a second
time.




We apply resource acquisition ordering to solve this problem.  Resource ordering
can be imposed by taking advantage of the fact that protocol layers and services
are composed in certain orders. Although packet routing is dynamic, we assume
that all possible paths through the network stack for supported packet types are
known when the network stack is compiled. As a simple example, Figure
\ref{fig:deadlock_exs} shows different topologies for the receive path of a UDP
stack. \thesystem's NoC uses wormhole, dimension-ordered routing. The packet
should be processed by Ethernet, IP, UDP, and then the application. With the
tile layout in Figure \ref{fig:simple_deadlock}, the route from the Ethernet to
IP tile passes through the UDP tile's router (2). As the UDP tile attempts to pass
the packet along to the application (4), it must reacquire a NoC link that is still in
use (5) and is thus deadlocked. 
If tiles are laid out as in Figure \ref{fig:simple_nondeadlock}, no resources
need to be reacquired, and the packet can be processed successfully.

We statically analyze all message paths in our prototypes at compile-time to avoid
deadlock by creating a resource dependency graph that takes into account every
possible path through the network stack. If a message path is found that could
cause deadlock, the designer should modify the tile layout to one that does not.

Repeated protocol headers (e.g. two IP headers in the IP-in-IP protocol) break
resource ordering.  In \thesystem, we choose to duplicate tiles (e.g. two IP RX
tiles).  If tiles are too expensive to duplicate, a potential solution is adding
buffers to break dependencies
\cite{message_deadlock_recovery_avoidance_lankes_nocs,mdisha_song}.  These
buffers give space for the NoC to drain into, freeing routing resources.



\subsection{Control plane interfaces}\label{sec:beehive_ctrl_plane_design}
For manageability, network operators need to be able to reconfigure protocol
components from an external controller over a transport-layer connection. In
\thesystem, we choose to use an additional separate message-based, routed NoC for the control plane rather than a dedicated control bus.
This is because control plane management also benefits from a structured interconnect for scalability reasons.

First, for complex designs with a large number of components, it becomes costly
to run dedicated, ad-hoc wires to every tile. Second, we want configuration to
be over a reliable transport. This requires the control plane to use the
transport layer, and a NoC enables this without physically coupling the
component to the transport layer. This also enables us to add specific control
plane management tiles to orchestrate state modifications. We describe a
specific example in Section \ref{sec:beehive_impl_nw_function}.

Because the control plane has lower performance requirements, in \thesystem{} we
use a separate, lower-width NoC. This also prevents control plane traffic from
contending for the same resources as long dataplane chains in the deadlock
dependency graph, so there is more flexibility in placement.

\subsection{Application interfaces}
Many application accelerators process requests at a coarser granularity than a
packet, so they need the ability to communicate with the transport protocol
layer and request data from a particular flow rather than being pushed packets
in the order they arrive. While we could use dedicated wires for
this communication, it can also benefit from the use of the NoC. 

The NoC provides a convenient structure to multiplex between duplicated
application tiles connected to the same transport layer in a scalable manner.
The modularity provided by message passing on the NoC also allows an application
to easily interface with any protocol in the network stack while reusing
existing wires if, for example, we want to switch from TCP to a custom reliable
transport protocol. Finally, the standardized NoC interface enables easy
insertion of filters on the application's NoC messages, so network operators can
enforce policies, such as dropping network traffic to or from non-whitelisted
nodes. We describe the application NoC interface to our TCP layer in Section
\ref{sec:beehive_impl_tcp}.

\vspace{-.1cm}
\section{Implementation}\label{sec:beehive_implement}
To demonstrate the \thesystem{} approach, we built a set of core protocol tiles,
network functions, and applications. For protocols, we implement tiles for
Ethernet, IPv4, UDP, and TCP. For network functions, we implement an IPinIP
encapsulation layer and a NAT layer for network virtualization. For
applications, we implement a Reed-Solomon encoder and an accelerator for a
viewstamped replication node. These applications are described in more detail in
Section \ref{sec:beehive_integration}.

We also describe our tooling that we developed to lower the effort required to
maintain multiple designs and integrate new components. All of \thesystem{} is
implemented in standard SystemVerilog and was tested on an Alveo U200
communicating with standard CPU clients using a Linux or kernel-bypass network
stack. We embed our \thesystem prototype within Corundum \cite{corundum_paper},
an open-source 100 Gbps NIC, in the application slot to provide FPGA-specific
infrastructure, such as the Ethernet MAC. Corundum does not provide any
higher-level packet processing logic for \thesystem. 

\subsection{Network-on-chip (NoC)} 
We use the 2D mesh NoC from OpenPiton \cite{openpiton_asplos} with some modifications. 
The NoC is wormhole-routed, uses dimension-ordered routing, and
is full-duplex. 
We widen the NoC from 64 bits to 512 bits to match the width of the Xilinx MAC IP core, so it has a maximum throughput of 128 Gbps when running at 250 MHz and increase the flit width to 512 bits.
Because the NoC only relies on the top 64 bits of the first flit to do NoC routing, we are able to reuse the NoC without further modification by making the top 64 bits of our first header flit the same as the original NoC header.
The maximum payload size for a NoC message is 256\,MiB.
\subsection{Protocol tiles} 
Protocols are implemented as streaming components, so they begin to transmit the next NoC message as soon as possible rather than storing the entire NoC message before forwarding. 
This is done to reduce latency as header parsing can be overlapped with payload copying. 
This is especially important when chaining, because each layer of header adds an extra layer of parsing. 

The Ethernet, IP, and UDP tiles construct or remove the appropriate headers and
calculate checksums, as shown in Figure~\ref{fig:beehive_basic_ex}. The Ethernet
receive processor can handle VLAN tagged packets. Our IP layer does not support
IP fragmentation as our intended use case is for internal datacenter services. 

One of the more difficult aspects of removing the headers from network packets is that certain protocols (e.g. IP or TCP) allow headers to have options, so the headers are not a fixed width. This means removing a packet header often requires removing a variable number of bytes from the stream.
We implement this by appending two lines of data and then using a shifter to remove the required amounts of bytes.

For a protocol, we place the receive and transmit engines in separate tiles. 
This is because they are streaming and each router has one input and one output interface, so one engine will utilize an entire router's bandwidth if running at 100 Gbps.
Since the packet-level protocol layers do not share state between
their transmit and receive sides, this is a straightforward split.
The exception to this is the TCP engine which we discuss further in Section~\ref{sec:beehive_impl_tcp}.

Protocol tiles also have optional hash tables that use the 4 tuple as the key
for load balancing to downstream replicated tiles. We set up initial
packet-level routing within the tiles at compile time when we build the FPGA
image. The hash table can be rewritten during runtime via the control plane described in Section~\ref{sec:beehive_ctrl_plane_design}. Any packet that does not
have an entry for a next hop (e.g. traffic with an unsupported protocol) is
dropped to filter out unwanted traffic.

\subsection{Buffer tiles}
In \thesystem, we also have buffer tiles that hold large blocks of memory. In
our current prototype, these buffers are large BRAMs, but the backing buffer
can also be DRAM. These buffer tiles are accessible to any other tile
in the system via NoC messages. This allows us to have shared buffers between
tiles, so that multiple tiles can share state when needed.



\subsection{TCP engine}\label{sec:beehive_impl_tcp}
To evaluate how \thesystem can support reliable transport, we prototype
a TCP engine that implements server-side TCP. It can receive connection setup
requests, generate sequence and ACK numbers, and support fast retransmit and
window-based flow control \cite{rfc_fast_retransmit}. Currently, it does not
support selective acknowledgments, initiating connections, or congestion
control. Full TCP offload functionality has been demonstrated by previous work
\cite{limago} and could be integrated into \thesystem.

We split the TCP logic into receive and transmit engines.
The receive engine is responsible for determining if received data is in order, calculating the next
ACK, and processing ACKs for the transmitted data. 
The transmit engine is responsible for separating out buffers for sending and updating the sequence
number for the transmitted stream. 

We use two optimizations to handle state shared between the receive and transmit engines when they are both processing the same flow.
We handle this in two ways.
First, we divide flow state into two BRAMs by which engine writes the data to prevent write conflicts. 
Second, we take advantage of the asynchronous nature of the transmit and receive streams in TCP to tolerate slightly stale state and avoid bypassing state when the two engines are processing the same flow.
For example, the transmit engine reads the current flow state with the ACK number for the received stream as $ACK\_RECV_1$ in cycle $n$.
Meanwhile, the receive engine has processed a packet and updated the ACK number to $ACK\_RECV_2$ in cycle $n+1$. 
The transmit engine can still use $ACK\_RECV_1$ as long as it still uses all the other state it read in cycle $n$. 
Functionally, this is the same as if the received packet had been received slightly later and processed after the transmit engine had sent its packet, which is allowed due to the assumptions of TCP.

While the TCP engine has an RX router and a TX router like the other protocol
tiles, the send and receive paths in TCP must share state.  For example, the
transmit path needs to know for which packets it has received acknowledgments.
We choose to support sharing by running dedicated wires between the tiles. Every
receive path only has one corresponding transmit path, so wires do not fan out. We could implement state sharing over NoC messages, but the state is read and updated frequently, so the frequent NoC
messages needed for state updates would encourage these tiles to be placed close to each other on the NoC anyway.

On the completion of the 3-way handshake, the TCP engine sends a NoC message to
notify an application tile based on the destination port for the connection. On
the receive side, the TCP engine lets an
application specify the size of the request it should be notified for with a NoC
message. When enough data has arrived to satisfy that request, the TCP engine
sends a notification message back to the application with the buffer address
where the data requested has been stored. The application then retrieves the
data from the buffer for processing before sending another message to the TCP
stack when it has finished using the data. 

The TCP engine implements a similar interface for the transmit engine where the
application can request space in its transmit buffer of a certain size. The TCP
engine sends a notification when there is room in that buffer with the 
address where the data should be stored. The application then copies the data
into the buffer and notifies the TCP engine.

\subsection{Network function tiles}\label{sec:beehive_impl_nw_function} We
implement both IPinIP encapsulation and an IP NAT. 
For both tiles, the control plane can dynamically update the table mapping virtual IPs to physical IPs, which occurs when the a client machine migrates. 
To change this mapping, we implement an internal controller as a separate tile that receives an RPC over TCP from an external controller.
The internal controller utilizes the control NoC to send NoC messages to the IP encapsulation or NAT tiles with the information needed to update their tables.
Finally, the internal controller sends a confirmation response to the external controller.

\subsection{Debugging and logging}
In \thesystem, tiles may keep logs, and we provide UDP and TCP-based protocols to externally fetch logs. 
Each log is associated with a particular port and exposes an interface on the NoC to the network stack for readback.
The layer 4 receive tiles are responsible for directing packets to the appropriate log interfaces.
The log read interface keeps a small buffer for requests and drops requests when it is full.
The client program reads out the log an entry at a time and resend requests for any entries for which it does not receive a response. 

This logging ability was invaluable for debugging TCP when running on an FPGA. 
TCP is underspecified and the main verification is running against a common implementation, such as the Linux kernel~\cite{testing_tcp_cam}, so we needed to run it on an FPGA to verify that it behaves as expected. The reduced visibility in this setting increases the difficulty of the already hard task of debugging a TCP implementation, due to the asynchronous and non-deterministic setting where certain bugs are dependent on the available bandwidth and loss events. 
As a result of the asynchrony, we need a cycle accurate trace for proper replay, because the TCP engine may behave differently depending on the timing of events (e.g. it may drop different packets). 
As a result of the bandwidth-dependence, we cannot rely on \texttt{tcpdump} to collect traces, because of the possibility different packets might be dropped by the engine versus \texttt{tcpdump}.

We inserted tiles that log information about TCP packet headers into the processing between the TCP and IP layers.
These tiles have two NoC interfaces: 
one is used to forward packets to and from the TCP engine and logs the header information with a cycle timestamp, 
the other interface allows the logs to be read out over the network in response to a request sent over UDP. 
Because the logging tiles are embedded within the fabric, they can record the exact timing and sequence of packets that entered and exited the TCP engine. 
Once this log is collected, we are able to replay the log in a cycle-accurate manner using the recorded timestamps by replacing the logging tiles with an interface to our trace replay framework.



\subsection{Tooling}
\vspace{-.1cm}
We developed a set of tools to lower the engineering effort to create new
designs, such as generating portions of the Verilog (e.g. top-level wiring for
NoCs) or performing compile-time deadlock analysis. The design configuration is
passed to these tools via an XML file, which contains the design dimensions as
well as an element for each NoC tile endpoint.  At minimum, this element
contains tags specifying a name to use for the endpoint as well as its X and Y
coordinates. It may also contain fields with information for generating the
tables used for determining the correct next hops. 

Given the dimensions in the XML file, we generate declarations of all the
top-level wires between tiles. We also generate the subset of the port
connections for each tile that correspond to wires between NoC routers and connect
the appropriate wires for the tile configuration. We choose not to generate the
whole tile instantiation, because certain tiles need to maintain additional
ports for I/O, such as the Ethernet MAC.

\begin{figure}[t]
    \centering
    \includegraphics[clip=true, trim={0.7in 0.125in 2.625in 0}, width=.9\linewidth]{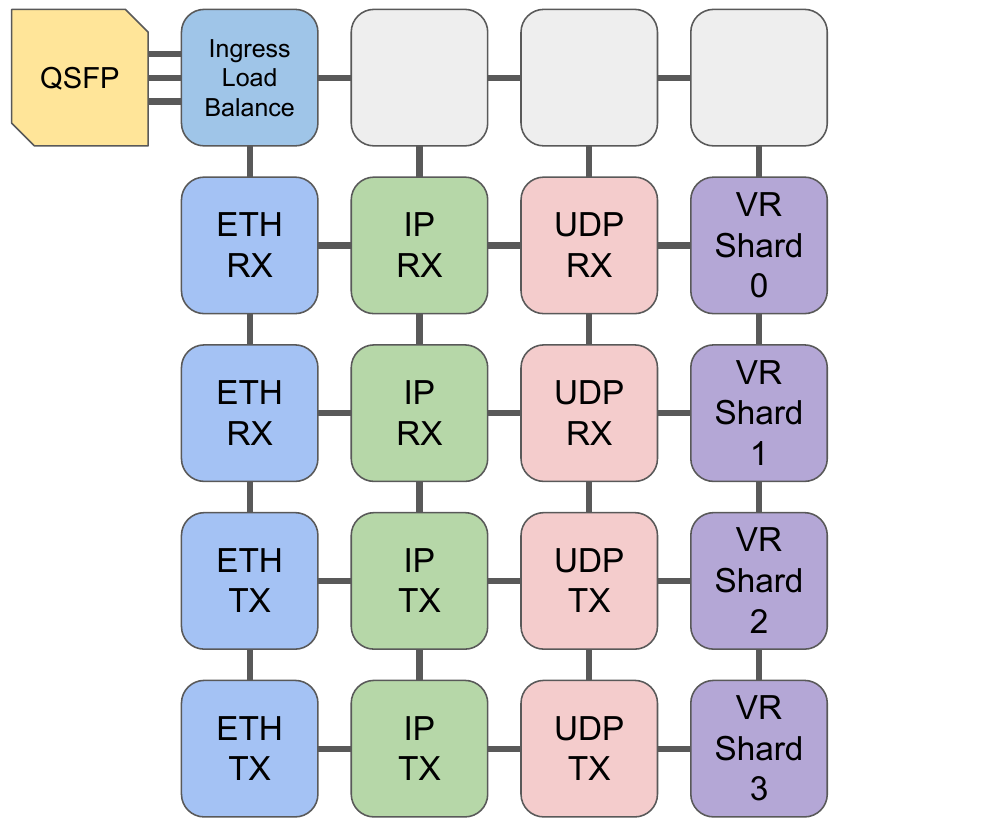}
    \caption{\thesystem{} tile layout for Viewstamped Replication.}
    \label{fig:beehive_vr_tiles}
\end{figure}

The XML file also enables us to check whether the high-level topology of the NoC is sound.
For example, we check if two tiles have the same X and Y coordinates, and all NoC coordinates are within the expected dimensions of the design.
Because a 2D mesh must be a rectangle, this also gives us the opportunity to automatically generate empty tiles that just contain a router, as in the bottom rightmost tile in the UDP stack shown in Figure \ref{fig:micro_beehive_setup}. 
We also use information about the NoC topology and next hops in the XML file to generate a resource dependency graph that we analyze for cycles to ensure a deadlock-free design. 
Figure~\ref{fig:beehive_vr_tiles} is a visualization of the layout generated by the XML file for the consensus witness design in Section~\ref{sec:consensus_witness}.

\section{Integrating with \thesystem}\label{sec:beehive_integration}

\subsection{Erasure coding}
To demonstrate the benefits of \thesystem for a throughput-oriented application, we integrate an accelerator for Reed-Solomon encoding.
Erasure codes such as Reed-Solomon (RS) are commonly used in distributed storage systems to achieve high resilience to disk failures with modest storage overhead~\cite{azureerasure,metaerasure,fasterasure}.
An RS encoder adds redundancy bits to input data at a pre-set ratio, striped across storage servers.
If some storage elements fail, the remaining blocks from the stripe can be combined with these extra blocks to regenerate the missing blocks.

We configure our system to use an (8,2) code (8 data blocks and 2 redundancy
blocks) to emulate a storage system that could tolerate up to two disk failures.
We integrate an RS encoding accelerator operating on 4KB requests into
\thesystem as a UDP application, instantiating four copies of the application to
scale out.  The accelerator is stateless, so any request can go to any copy. We
introduce a front-end round-robin scheduler tile to distribute work among the RS
tiles.  Each RS tile also logs metadata to calculate bandwidth.


\subsection{Consensus witness}\label{sec:consensus_witness}
To demonstrate how \thesystem performs in a  latency-sensitive, communication-intensive application, we construct a consensus system that uses FPGA-accelerated witness nodes.
Consensus algorithms are an essential part of many deployed distributed systems as they enable a strictly consistent order for stateful client operations even in the face of failures and message delays/retransmissions.
Most consensus algorithms~\cite{liskov12vr,paxoslive,raft} follow a common pattern: an elected leader proposes an order for arriving client requests, verifies with a set of replicas that it is still leader, and commits the request.
It then performs any necessary application logic (e.g., to update state), replies back to the client, and informs the other replicas, so that they can also perform the application logic in the same order.
Because there are multiple round-trips between nodes to complete one round of consensus, message-handling latency and tail latency are especially important~\cite{electrode_nsdi23}.

A common type of application built on top of consensus is a key-value (KV)
store. To achieve higher throughput, the key space is often sharded with a
leader and replica set for each slice. However, even with sharding, consistent
reads can be expensive, because the leader must validate, each time, that it is
still the leader before replying with the value stored with the key. As a
result, it is common in practice to configure the system to return stale reads,
allowing the leader to reply
immediately~\cite{zookeeper,crooksweak,caribou_vldb_eth}. This places a burden
on the client developer to handle the (rare) case where a failover can lead to
inconsistent client data.

In our evaluation, we show that a consensus accelerator can help reduce the cost of consistency~\cite{istvan2016consensus}, especially in a multi-shard setting.
Our accelerator operates as a witness, that is, it only validates the leader and tracks the operation order; it does not execute client operations.  
Single node fault tolerance can be achieved with one leader, one witness, and one replica.
To add further fault tolerance, we add additional witnesses and replicas.
For example, two-node fault tolerance can be achieved with one leader, two
witnesses, and two replicas. 
To validate a read or write operation, the leader
only needs to receive a verification from the witnesses before replying to the
client.
The witness can be designed in hardware to reply with low and reliable
latency.

Prior work \cite{istvan2016consensus,caribou_vldb_eth} has demonstrated full offload of consensus and application logic to an FPGA.
We target a use case where application logic remains on CPUs and only a portion of the consensus protocol is run on
\thesystem.  Importantly, this requires no change to the CPU-based application running on top of the consensus engine.
This is advantageous as consensus algorithms are commonly used as a building block in larger distributed systems, so this allows accelerated consensus to be used without requiring the whole application to be ported to hardware.
We also demonstrate how \thesystem can be used to scale a consensus system to support multiple shards, which previous work did not explore.

Our witness protocol is based on a modified version of the Viewstamped
Replication (VR) used in previous studies of high-performance
consensus~\cite{specpaxos}. VR witnesses are integrated into \thesystem as UDP
applications.  To handle multiple shards, we use one VR witness
tile per shard. Unlike the RS encoder, the VR witness is not stateless and requests for a
shard must always go to the same tile. We distribute work to the VR tiles by
matching on the destination port number.

%

\section{Evaluation}
Our evaluation tests \thesystem's ability to support scalability, low latency,
and flexibility in a range of network stack configurations. 
We begin by evaluating \thesystem{} with UDP and TCP microbenchmarks designed to test RPC performance and then evaluate two case studies:
Reed-Solomon encoding acceleration and Viewstamped Replication acceleration.


\vspace{-.1cm}
\subsection{Setup}
We use Vivado 2021.2 for building our FPGA images. \thesystem is configured on
an Alveo U200 at 250 MHz. The FPGA and the clients are connected to an Arista
DCS-7060CX-32S-R 100G switch with jumbo frames enabled. We use five machines
during evaluation with Turboboost disabled. All of them have Mellanox ConnectX-5
100G NICs and are running Ubuntu 20.04. Two machines have Intel Xeon Gold 6226R
CPUs; the other three machines have Intel Xeon Gold 5218 CPUs.

In experiments where energy is measured, we use the RAPL counters on the CPUs
and the Alveo CMS registers on the FPGA. For CPU energy experiments we use a
two-socket machine, so we run all the application and network processing code on
one socket and poll the counters from the other socket. We only use RAPL's CPU
counters, which is an underestimate as we do not include DRAM energy or network
interface energy. On the FPGA, we use the Corundum framework to read the CMS
registers that report instantaneous power and current
usage~\cite{alveo_cms_register_space}. We poll these counters every second to
calculate energy over the benchmarking period. 

\subsection{Baselines}

\noindent\textbf{Hardware Network Stacks (PANIC and Limago):} We compare against
PANIC, an FPGA-based smartNIC framework, for our UDP echo
microbenchmark. We are unable to compare against PANIC for our other
applications using UDP, because they require scaling to more tiles than PANIC
supports, and PANIC's memory allocation makes it unwieldy to generate responses
of a different size than the request. We also cannot compare against PANIC for
our TCP microbenchmark, because it cannot support reliable transport
applications.
We also evaluate in PANIC's original simulation evaluation infrastructure, because their released code does not include an FPGA flow. 
We integrated it into Corundum as suggested in the documentation, but we were unable to get it to meet timing for the Alveo U200.
While they used an ADM-PCIE-9V3~\cite{panic_fpga_datasheet}, both our board and theirs have 16nm FPGA parts. The Alveo's FPGA part is also comparable in resources available to the ADM-PCIE-9V3. For these reasons, we think the comparison is fair.

We compare against Limago, an HLS TCP stack, for our hardware utilization. We were unable to run benchmarks on it, because the QSFP links did not come up when the image was put on the board.

\noindent\textbf{Software Network Stacks (Demikernel and Linux):} We also
compare against Demikernel~\cite{demikernel_sosp}, an optimized DPDK network
stack, in cases where it is faster than Linux. This is only the case in the UDP
echo benchmark. Otherwise, we compare against Linux's network stack.

\begin{figure}[t]
    \includegraphics[width=.95\linewidth]{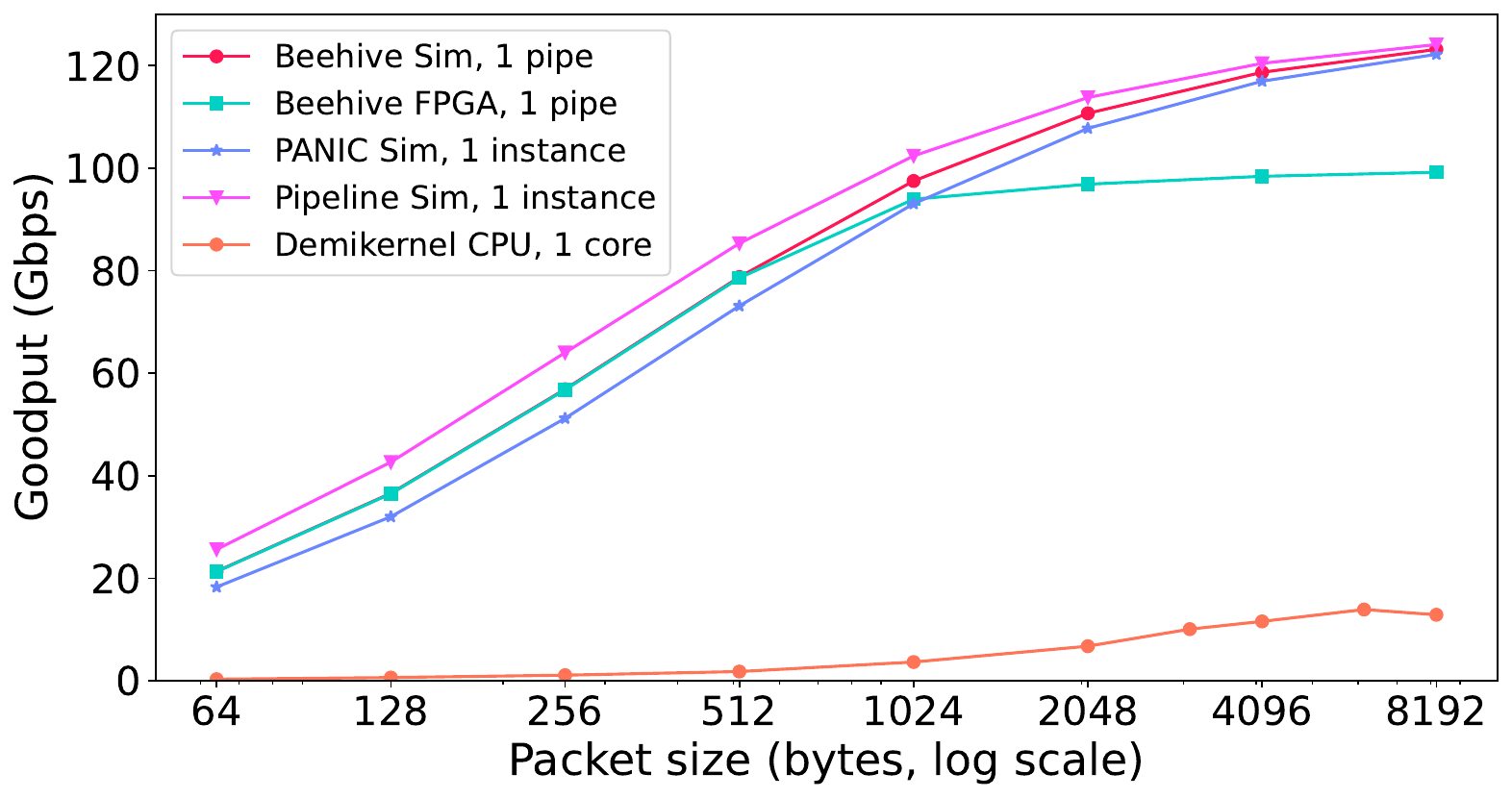}
    \caption{Packet size vs. goodput for a UDP echo application. \thesystem and
    \PANICUdpEcho perform almost identically across all packet sizes and
    outperform Demikernel.}
    \label{fig:udp_packet_size_vs_bw}
\end{figure}

\subsection{UDP echo}\label{sec:beehive_udp_microbench}

\begin{figure}
    \centering
    \begin{subfigure}{\linewidth}
        \includegraphics[width=\linewidth]{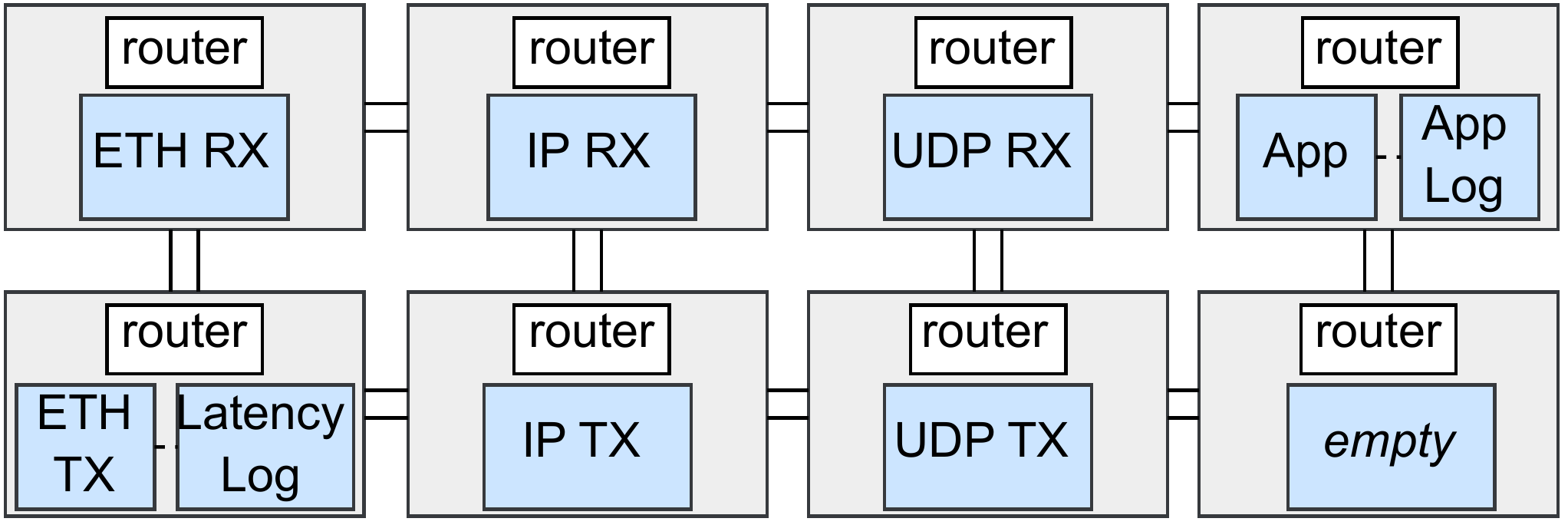}
        \vspace{-.5cm}
        \caption{Setup for \thesystem's UDP stack}
        \label{fig:micro_beehive_setup}
    \end{subfigure}
    \hfill{}
    \begin{subfigure}{\linewidth}
        \vspace{.2cm}
        \includegraphics[width=\linewidth]{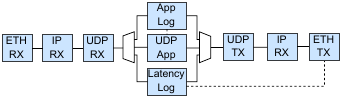}
        \caption{Setup for the fixed pipeline UDP stack}
        \label{fig:micro_fixed_pipe}
    \end{subfigure}
    \caption{UDP stacks for the echo microbenchmark}
\end{figure}

\noindent\textbf{Throughput:} We compare UDP echo goodput for \thesystem (shown in Figure~\ref{fig:micro_beehive_setup}) and Demikernel on different packet sizes. 
We also evaluate these against an FPGA with a pipelined UDP network stack design where the protocol engines are connected directly (shown in Figure~\ref{fig:micro_fixed_pipe}), and a UDP network stack implemented within the PANIC framework which we will refer to as \PANICUdpEcho.

In our experiments, the Demikernel server runs on an Intel Gold 6226R
machine, and we use three Intel Gold 5128 machines as clients using the standard
Linux network stack. We spawn the number of client threads that yields the
highest server bandwidth for that packet size, and they send in an open-loop
manner. We give the server a single core to compare against \thesystem's single application tile.

For \thesystem, we run a packet generator on another U200 FPGA. This is
because the client machines used for the CPU experiments cannot generate enough
traffic to saturate the FPGA. We use 7 tiles in total: we separate the Ethernet,
IP, and UDP layers, and then we separate their receive and transmit paths for 6
tiles and then one tile for the application.

For \PANICUdpEcho, we implement a UDP echo server within its framework
starting from their publicly available code~\cite{panic_bitbucket}. 
We use 3 tiles to implement the echo server: one providing a fixed UDP receive path, one providing the application, and one providing a fixed UDP send path.
We were unable to modify PANIC to support more than 8 tiles, only 4 of which are
available for user functionality, so we could not make every layer into a tile
as we do in \thesystem. 
We note that this means it is less flexible than \thesystem's network stack, because we lose the opportunity to easily insert network functions or alternate protocols alongside the UDP paths. 

Figure \ref{fig:udp_packet_size_vs_bw} shows our throughput benchmark
results. 
\thesystem and \PANICUdpEcho provide similar performance despite \thesystem having more tiles. 
Both achieve line rate at 1024 byte packets.
\thesystem on FPGA levels out at this point, because the actual Ethernet link has a maximum bandwidth of 100 Gbps. 
However, in simulation, both \thesystem and \PANICUdpEcho continue to scale to the theoretical maximum of 128 Gbps. 
The pipelined implementation is slightly better than \thesystem, due to the overhead of constructing and deconstructing NoC messages. However, this difference decreases as payload sizes increase since the extra header flits are amortized over a larger payload. 
The optimized CPU stack remains far below maximum bandwidth even with jumbo frames.
The performance difference is especially pronounced at small packet sizes where \thesystem is able to sustain echoing 9\,Gbps of 64-byte packets (18392\,KReq/s)
whereas single core Demikernel provides 0.3\,Gbps (584\,KReq/s), a 31$\times$
speedup.

\noindent\textbf{Latency:}\label{sec:micro_latency}
For our latency experiment, we use \thesystem{} and a single
client thread to ping-pong a single 1-byte UDP packet. We record the latency by
tagging the packet with a timestamp when it enters the network stack at the
Ethernet parsing layer, taking another timestamp when it finally exits the
Ethernet layer on transmit, and recording both timestamps into a log which we read
back over the network. The latency through \thesystem is 368\,ns (92 cycles).
Similarly, \PANICUdpEcho UDP latency is 362\,ns, although their system is less flexible than \thesystem.


\begin{figure}[t]
    \centering
    \includegraphics[width=.95\linewidth]{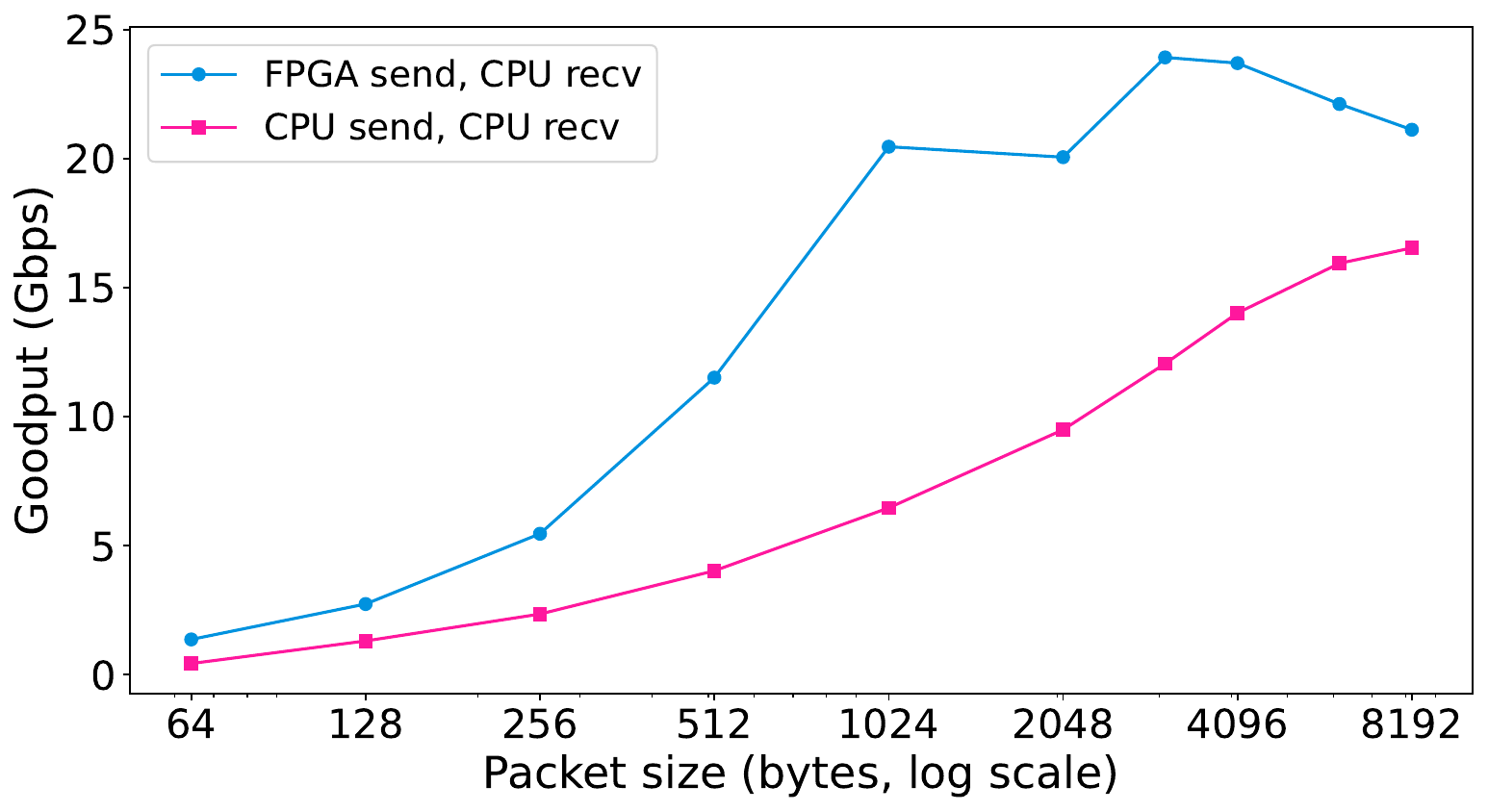}
    \caption{Packet size vs. goodput for \thesystem and Linux TCP send. The (CPU
    send/FPGA receive) is omitted, as it is approximately the same as (CPU
    send/CPU receive) due to the CPU send path being the bottleneck.}
    \label{fig:tcp_goodput}
    \vspace{-.2cm}
\end{figure}

\def\thickhline{\noalign{\hrule height.7pt}}

\begin{table}[t]
    \centering
    \caption{Energy consumption and goodput for Reed-Solomon encoding using
    \thesystem versus CPU for 1, 2, 3 and 4 application instances.}
    \resizebox{\linewidth}{!}{
    \begin{tabular}{rllll}
        Apps & 1 & 2 & 3 & 4\\
    \toprule
       CPU Energy (mJ/op) & 1.1  & 0.59 & 0.41 & 0.32 \\
        \thesystem Energy (mJ/op) & 0.05 & 0.03 & 0.02 & 0.02 \\
\cline{2-5}\noalign{\vspace{0.4ex}}
Energy efficiency      & 22$\times$ & 20$\times$ & 20$\times$ & 16$\times$ \\ 
\midrule
CPU Goodput (Gbps)     & 2.0  & 4.0  & 6.0  & 8.0  \\
\thesystem Goodput (Gbps) & 15  & 31  & 45  & 62  \\
\cline{2-5}\noalign{\vspace{0.4ex}}
Speedup         & 7.5$\times$ & 7.8$\times$ & 7.5$\times$ & 7.8$\times$ \\
\bottomrule
    \label{tab:rs_encode_scaling}
    \end{tabular}
    }
\end{table}

\subsection{TCP throughput}
To characterize the throughput performance of our TCP engine, we run a
single-connection experiment and measure unidirectional send and receive performance across a range of packet payload sizes.
Because Demikernel's TCP implementation is optimized for latency, it performs worse than Linux on this experiment, so we configure Demikernel to use Linux TCP as its backend.
The sending application sits in a tight loop, submitting data into the network stack as fast as possible; the receiver pulls data out of the network stack without doing further processing on it.

We vary whether the sender or the receiver is the FPGA or the CPU.  The results
are shown in Figure \ref{fig:tcp_goodput}. We omit the (CPU send/FPGA receive)
results, because they are almost the same as the all-CPU configuration; in both
situations, the CPU sender is the bottleneck. The CPU is more efficient at
streaming TCP data than UDP data because it allows batching data into jumbo
frames. By contrast, \thesystem's TCP stack is slower than its UDP stack, because
of the complexity of stateful packet handling in hardware. In particular, our
TCP engine is designed to only achieve full bandwidth across multiple
simultaneous connections. Even so, \thesystem outperforms Linux TCP across all
request sizes. The speedup is most pronounced at small packet sizes, where
\thesystem achieves 2666\,KReq/s versus the CPU's 843\,KReq/s, a 3.2$\times$
speedup.

\subsection{Reed-Solomon encoding acceleration}
To evaluate \thesystem's scaling architecture, we evaluate a duplicated
Reed-Solomon (RS) encoding accelerator on \thesystem versus a CPU implementation
of the same algorithm in Table \ref{tab:rs_encode_scaling}.
The client sends blocks of 4\,KB to the encoder using UDP;
the accelerator replies with 1K of erasure data. This could be organized into an
(8,2) stripe for double fault tolerance. We measured that one instance of the Reed-Solomon
encoder can consume data at 15\,Gbps; our FPGA has room for four encoder
instances, which consume data at 62\,Gbps as shown in Table
\ref{tab:rs_encode_scaling}. For comparison, we use the open-source Reed-Solomon
encoding implementation from BackBlaze~\cite{backblaze_rs_encode} running on
CPUs which we then duplicate across cores. 

We also compare the energy efficiency of the two approaches in Table
\ref{tab:rs_encode_scaling}. The FPGA is about 20$\times$ more efficient per operation
than the CPU implementation.


\subsection{Viewstamped replication witness acceleration}
\begin{figure}[t]
    \centering
    \begin{subfigure}[t]{.49\linewidth}
        \includegraphics[width=\linewidth]{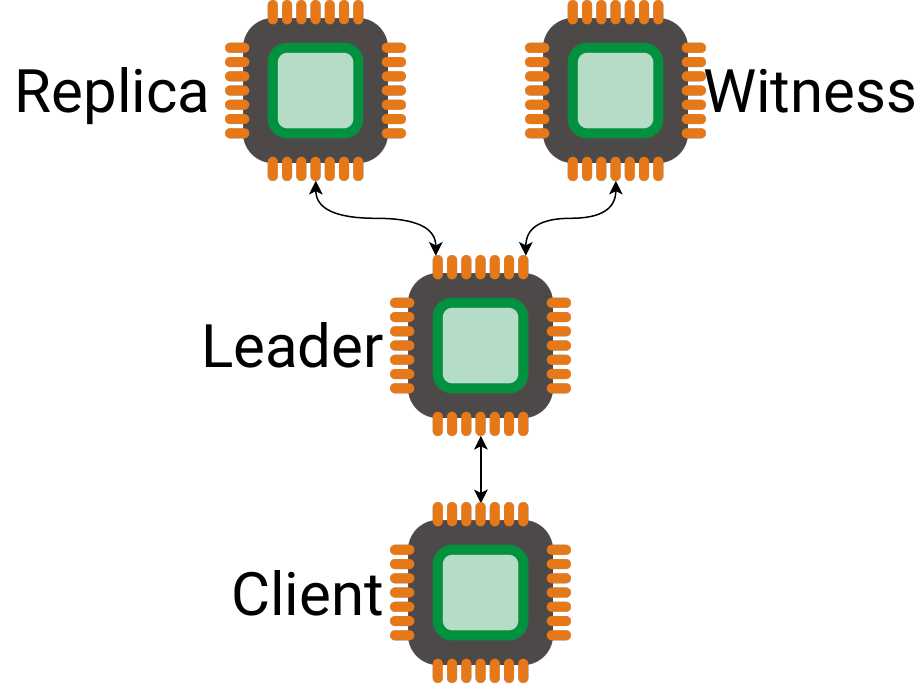}
        \caption{CPU-only setup} 
    \end{subfigure}
    \hfill
    \begin{subfigure}[t]{0.49\linewidth}
        \includegraphics[width=\linewidth]{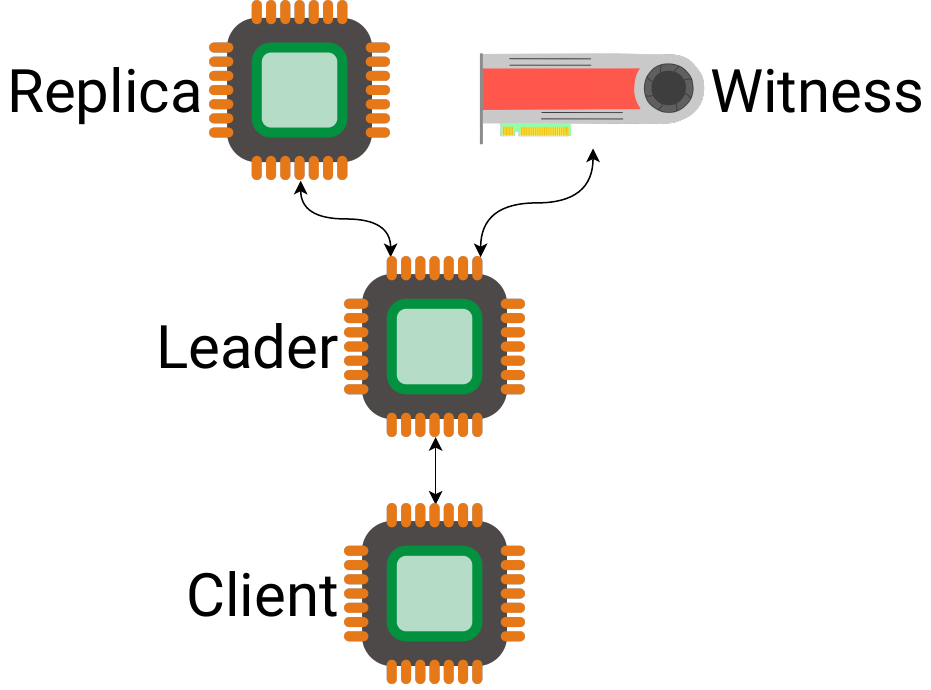}
        \caption{\thesystem witness setup}
    \end{subfigure}
    \caption{Experimental setups for VR evaluation}
    \label{fig:vr_experiment_setup}
    \vspace{-.5cm}
\end{figure}
We next turn to a latency-sensitive application, evaluating \thesystem{} hosting
a viewstamped replication (VR) witness appliance.  We first evaluate the witness
on a single shard. We then take advantage of \thesystem's ability to duplicate
both internal components and applications to host a 4-shard witness appliance.
We also duplicate protocol tiles to prevent them from becoming a bottleneck.

\noindent\textbf{Setup:} For all experiments, we evaluate a three-node VR
configuration as shown in Figure \ref{fig:vr_experiment_setup}, with either the
FPGA or CPU serving as a witness. Other nodes are run on CPUs. The CPU VR
replicas run on Intel Xeon Gold 5218 CPUs. Client threads run on Intel Xeon Gold
6226R CPUs and are closed loop, i.e., only have one outstanding request at a
time. The shard leaders are distributed evenly between two CPU machines. Each
shard may handle more than one request at a time. The CPU witness(es) run on a
separate server to allow us to measure the energy used by a CPU witness
appliance. We use UDP as our transport protocol, because VR does not assume
reliable message delivery.

%
%

\noindent\textbf{Workload and Metrics:} We evaluate our VR accelerator with a
replicated key-value store application with 64-byte keys and 64-byte values. The
workload uses a read-write mix of 90\% reads and 10\% writes and a uniform key
distribution. Input load is increased by increasing the number of clients.
Latency is measured at the clients as the time between the initial request and
the eventual response. Peak throughput numbers are chosen at the points before
the latency begins to spike, an indication that the system is overloading and
queues in the system are growing. These points correspond to operational setups
where increased latency might be considered acceptable in exchange for better
throughput~\cite{lu2020performance,chawathe2003making,mills1991internet}.

\begin{figure}
    \includegraphics[width=\linewidth]{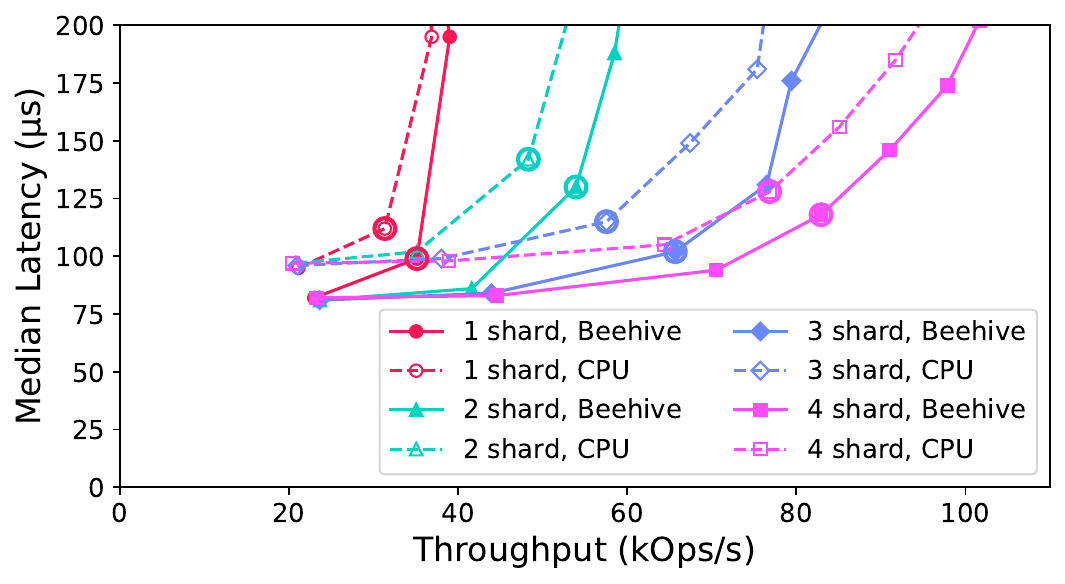}
    \caption{Latency vs. throughput for the VR key-value store workload varying the
    number of shards and client threads. The FPGA witness consistently
    outperforms the equivalent CPU cores in both latency and throughput.
    }
    \label{fig:vr_latency_v_thruput}
    \vspace{-0.2cm}
\end{figure}

\noindent\textbf{Results:}
We plot latency versus throughput for differing numbers of shards in Figure
\ref{fig:vr_latency_v_thruput}. We increase offered load by increasing the
number of client threads sending requests to the leader. The results are shown
in Figure~\ref{fig:vr_latency_v_thruput}. The system using the FPGA witness can
provide up to \VRthruputComparison{} more per-core throughput and up to
\VRlatencyComparison{} lower median latency.

%
%

\newcommand{\VRMedianLatencyCPUIShard}{112\xspace}
\newcommand{\VRMedianLatencyBeehiveIShard}{99\xspace}
\newcommand{\VRMedianLatencyBothIShard}{1.13$\times$\xspace}
\newcommand{\VRNNLatencyCPUIShard}{273\xspace}
\newcommand{\VRNNLatencyBeehiveIShard}{281\xspace}
\newcommand{\VRNNLatencyBothIShard}{0.97$\times$\xspace}
\newcommand{\VREnergyCPUIShard}{1.51\xspace}
\newcommand{\VREnergyBeehiveIShard}{0.73\xspace}
\newcommand{\VREnergyBothIShard}{2.07$\times$\xspace}
\newcommand{\VRThroughputCPUIShard}{31\xspace}
\newcommand{\VRThroughputBeehiveIShard}{35\xspace}
\newcommand{\VRThroughputBothIShard}{1.12$\times$\xspace}

\newcommand{\VRMedianLatencyCPUIIShard}{142\xspace}
\newcommand{\VRMedianLatencyBeehiveIIShard}{130\xspace}
\newcommand{\VRMedianLatencyBothIIShard}{1.09$\times$\xspace}
\newcommand{\VRNNLatencyCPUIIShard}{372\xspace}
\newcommand{\VRNNLatencyBeehiveIIShard}{334\xspace}
\newcommand{\VRNNLatencyBothIIShard}{1.11$\times$\xspace}
\newcommand{\VREnergyCPUIIShard}{1.03\xspace}
\newcommand{\VREnergyBeehiveIIShard}{0.48\xspace}
\newcommand{\VREnergyBothIIShard}{2.16$\times$\xspace}
\newcommand{\VRThroughputCPUIIShard}{48\xspace}
\newcommand{\VRThroughputBeehiveIIShard}{54\xspace}
\newcommand{\VRThroughputBothIIShard}{1.12$\times$\xspace}

\newcommand{\VRMedianLatencyCPUIIIShard}{115\xspace}
\newcommand{\VRMedianLatencyBeehiveIIIShard}{102\xspace}
\newcommand{\VRMedianLatencyBothIIIShard}{1.13$\times$\xspace}
\newcommand{\VRNNLatencyCPUIIIShard}{339\xspace}
\newcommand{\VRNNLatencyBeehiveIIIShard}{304\xspace}
\newcommand{\VRNNLatencyBothIIIShard}{1.12$\times$\xspace}
\newcommand{\VREnergyCPUIIIShard}{0.90\xspace}
\newcommand{\VREnergyBeehiveIIIShard}{0.39\xspace}
\newcommand{\VREnergyBothIIIShard}{2.32$\times$\xspace}
\newcommand{\VRThroughputCPUIIIShard}{58\xspace}
\newcommand{\VRThroughputBeehiveIIIShard}{66\xspace}
\newcommand{\VRThroughputBothIIIShard}{1.14$\times$\xspace}

\newcommand{\VRMedianLatencyCPUIVShard}{128\xspace}
\newcommand{\VRMedianLatencyBeehiveIVShard}{118\xspace}
\newcommand{\VRMedianLatencyBothIVShard}{1.08$\times$\xspace}
\newcommand{\VRNNLatencyCPUIVShard}{412\xspace}
\newcommand{\VRNNLatencyBeehiveIVShard}{394\xspace}
\newcommand{\VRNNLatencyBothIVShard}{1.05$\times$\xspace}
\newcommand{\VREnergyCPUIVShard}{0.70\xspace}
\newcommand{\VREnergyBeehiveIVShard}{0.31\xspace}
\newcommand{\VREnergyBothIVShard}{2.27$\times$\xspace}
\newcommand{\VRThroughputCPUIVShard}{77\xspace}
\newcommand{\VRThroughputBeehiveIVShard}{83\xspace}
\newcommand{\VRThroughputBothIVShard}{1.08$\times$\xspace}

\begin{table}[t]
    \centering  
    \footnotesize
    \caption{
	    Energy per operation (measured at the witness) and performance metrics
	    (measured at the clients) at the circled points in 
    Figure~\ref{fig:vr_latency_v_thruput}.}
    \begin{tabular}{rllll}
Shards & 1 & 2 & 3 & 4 \\
\toprule
\multicolumn{1}{r}{CPU Energy (mJ/op)}     & \VREnergyCPUIShard & \VREnergyCPUIIShard & \VREnergyCPUIIIShard & \VREnergyCPUIVShard \\
\multicolumn{1}{r}{\thesystem Energy (mJ/op)} & \VREnergyBeehiveIShard & \VREnergyBeehiveIIShard & \VREnergyBeehiveIIIShard & \VREnergyBeehiveIVShard \\
\cline{2-5}\noalign{\vspace{0.4ex}}
\multicolumn{1}{r}{Energy efficiency}      & \VREnergyBothIShard & \VREnergyBothIIShard & \VREnergyBothIIIShard & \VREnergyBothIVShard \\
\midrule
\multicolumn{1}{r}{CPU Throughput (kOps/s)}     & \VRThroughputCPUIShard & \VRThroughputCPUIIShard & \VRThroughputCPUIIIShard & \VRThroughputCPUIVShard \\
\multicolumn{1}{r}{\thesystem Throughput (kOps/s)} & \VRThroughputBeehiveIShard & \VRThroughputBeehiveIIShard & \VRThroughputBeehiveIIIShard & \VRThroughputBeehiveIVShard \\
\cline{2-5}\noalign{\vspace{0.4ex}}
\multicolumn{1}{r}{Speedup}                & \VRThroughputBothIShard & \VRThroughputBothIIShard & \VRThroughputBothIIIShard & \VRThroughputBothIVShard \\
\midrule

\multicolumn{1}{r}{CPU Median Latency ($\mu$s)}     & \VRMedianLatencyCPUIShard & \VRMedianLatencyCPUIIShard & \VRMedianLatencyCPUIIIShard & \VRMedianLatencyCPUIVShard \\
\multicolumn{1}{r}{\thesystem Median Latency ($\mu$s)} & \VRMedianLatencyBeehiveIShard & \VRMedianLatencyBeehiveIIShard & \VRMedianLatencyBeehiveIIIShard & \VRMedianLatencyBeehiveIVShard \\
\cline{2-5}\noalign{\vspace{0.4ex}}
\multicolumn{1}{r}{Improvement}                & \VRMedianLatencyBothIShard & \VRMedianLatencyBothIIShard & \VRMedianLatencyBothIIIShard & \VRMedianLatencyBothIVShard \\
\midrule
\multicolumn{1}{r}{CPU p99 Latency ($\mu$s)}     & \VRNNLatencyCPUIShard & \VRNNLatencyCPUIIShard & \VRNNLatencyCPUIIIShard & \VRNNLatencyCPUIVShard \\
\multicolumn{1}{r}{\thesystem p99 Latency ($\mu$s)} & \VRNNLatencyBeehiveIShard & \VRNNLatencyBeehiveIIShard & \VRNNLatencyBeehiveIIIShard & \VRNNLatencyBeehiveIVShard \\
\cline{2-5}\noalign{\vspace{0.4ex}}
\multicolumn{1}{r}{Improvement}                & \VRNNLatencyBothIShard & \VRNNLatencyBothIIShard & \VRNNLatencyBothIIIShard & \VRNNLatencyBothIVShard \\

\bottomrule
\end{tabular}
    \label{tab:vr_energy_use}
\end{table}
 
For each shard, we take the median energy measurement, throughput, median
latency, and 99th-percentile (p99) latency at each circled point in Figure
\ref{fig:vr_latency_v_thruput}.  These results are shown in Table
\ref{tab:vr_energy_use}. The FPGA is between \VRenergyMin~and \VRenergyMax~more
energy efficient per operation compared to the CPU while providing better
overall throughput and latency to key-value store clients.

\subsection{Hardware resource utilization}
\label{sec:hw_util}
\begin{table}[t]
    \centering
    \footnotesize
    \caption{FPGA resource utilization of selected modules in \thesystem and Limago.}
    \resizebox{\columnwidth}{!}{    
        \begin{tabular}{lll}
        \toprule
           &  LUTs (\# / \% total)  & BRAM (\# / \% total) \\
        \midrule
        \thesystem UDP full               & 58540 / 4.95             & 41 / 1.90 \\
        UDP RX Tile                       & 10054 / 0.85             & 9.5 / 0.44\\
        \hspace{3mm} Router               & \hspace{3mm}5946 / 0.50  & \hspace{3mm}0 / 0 \\
        \hspace{3mm} NoC Message Parsing  & \hspace{3mm}897 / 0.07   & \hspace{3mm}0 / 0 \\
        \hspace{3mm} UDP RX Processing    & \hspace{3mm}2912 / 0.25  & \hspace{3mm}9.5 / 0.44 \\
        UDP TX Tile                       & 10128 / 0.86             & 9.5 / 0.44 \\
        \hspace{3mm} Router               & \hspace{3mm}5955 / 0.50  & \hspace{3mm}0 / 0 \\
        \hspace{3mm} NoC Message Parsing  & \hspace{3mm}658 / 0.06   & \hspace{3mm}0 / 0\\
        \hspace{3mm} UDP TX Processing    & \hspace{3mm}3105 / 0.26  & \hspace{3mm}9.5 / 0.44 \\
        \cmidrule{1-3}
        \thesystem TCP/UDP stack & 144491 / 12 & 84.5 / 4 \\
        \thesystem TCP Layer & 41677 / 3.5 & 25 / 1.1 \\
        \hspace{3mm}TCP RX Processing & \hspace{3mm}10304 / 0.87 & \hspace{3mm}9 / 0.4 \\
        \hspace{3mm}TCP RX Router & \hspace{3mm}8847 / 0.74 & \hspace{3mm}0 / 0\\
        \cmidrule{1-3}
        Limago TCP/UDP stack & 116948 / 9.9 & 155 / 7.2 \\
        Limago TCP Layer & 52134 / 4.4 & 99 / 4.6 \\

        \bottomrule
        \end{tabular}
    }
    \label{tab:fpga_resource_utilization}
\end{table}

The hardware utilization of the \thesystem{} infrastructure is shown in Table
\ref{tab:fpga_resource_utilization}. For the UDP stack used in Section
\ref{sec:beehive_udp_microbench}, \thesystem components use 4\% of the LUTs
available on the Alveo U200 and 2\% of the BRAMs. In a tile, a router uses
around 6000 LUTs, twice the size of the UDP processing. For comparison with a
more complex module, we include the utilization of the TCP receive path.

We also compare our resource utilization to that of Limago.
We find that our design is larger in terms of LUT usage, but smaller in terms of BRAM usage. Most of our usage comes from the routers rather than the protocol logic, indicating that there is a cost to our increased flexibility. However, in the context of total resources available on the FPGA, the extra logic cost is relatively small.

\subsection{Flexibility} 
As a quantitative proxy for flexibility, we count the lines of code (LoC)
required to insert an additional instance of an implemented service (network
function or application) into the design for our three designs.  Results are
shown in Table~\ref{tab:loc_new_tile}.

\begin{table}[t]
\centering
\caption{Lines of code per new tile instantiation in \thesystem for end-to-end applications.
XML configuration numbers are given as LoC for declaring the tile plus the LoC
to add it as a destination.}
 \resizebox{\linewidth}{!}{
\begin{tabular}{lll}
                                             & \multicolumn{2}{c}{Lines of Code}              \\ \cmidrule{2-3} 
                                             & XML Config.          & Verilog Top Level \\
                                             \toprule
Reed-Solomon            & 25 + 6                     & 13                \\
Viewstamped Replication & 18 + (6$\times$\# of UDP tiles) & 17                \\
\bottomrule
\end{tabular}
 }
 \label{tab:loc_new_tile}
 \vspace{-0.4cm}
\end{table}

\subsection{Scalability}
We did two experiments to evaluate the scalability of \thesystem: one bandwidth-oriented and one hardware resource oriented.
For the bandwidth-oriented experiment we repeated the UDP echo experiment in cycle-accurate simulation, duplicating the UDP stack and adding a simple load-balancing tile at the front that splits flows evenly between the stacks. 
The maximum goodput the load balancer can achieve is 32Gbps for 64-byte UDP packet since each takes 4 cycles to process at the load balancer: 3 for the NoC message and 1 recovery cycle. 
We hit the maximum possible goodput of the load-balancer of ~32Gbps for 64-byte packets.
With two stacks, at small packet sizes, we also roughly double the bandwidth as with one stack.
This performance difference decreases at larger payload sizes and both stacks converge to the maximum possible goodput of the network link.

To evaluate hardware resource usage scability, we duplicate
echo application tiles connected to a UDP stack. On the Alveo U200, we can place
22 application tiles and 28 tiles total. We are limited by timing rather than
resource utilization; the critical path is between NoC routers. Each router is
fairly expensive, because the 512-bit width of the bus results in a number of
high-fanout wires. This is exacerbated by the fact that the FPGA part in the
Alveo U200 is made up of several chiplets, and chiplet crossings add significant
delay. Several FPGAs \cite{achronix_noc,xilinx_versal} now support hardened NoC
resources and could improve the quality of results.

\begin{figure}[t]
    \centering
    \includegraphics[width=0.9\linewidth]{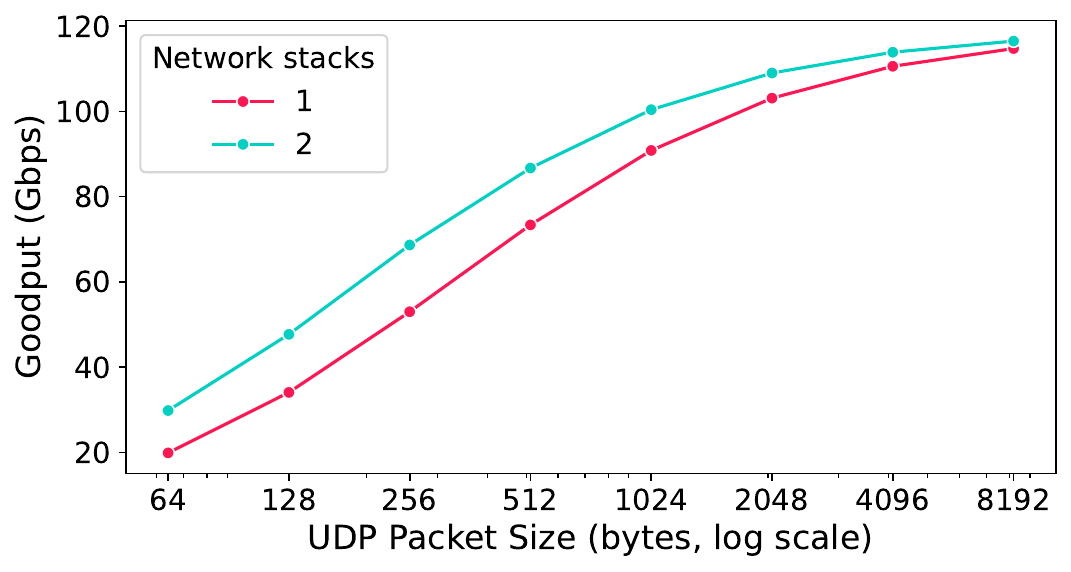}
	\caption{Packet size vs. goodput for a UDP echo application, running on \thesystem with multiple network stacks instantiated. 2 network stacks maxes out the load balancer's throughput.}
    \label{fig:udp_packet_size_vs_bw_scaling}
    \vspace{-.6cm}
\end{figure}

\section{Related Work}
\subsection{Packet processing}
PANIC~\cite{panic_osdi} is a smartNIC framework that supports integration of
arbitrary packet processing elements, including general purpose cores. Unlike
\thesystem{}, PANIC targets packet processing rather than full-stack support for
application accelerators.
PANIC uses a similar model to \thesystem{} of chaining message-passing
elements over a NoC, but it relies on a crossbar, limiting scalability.
While PANIC does not directly address deadlocks, its central scheduler drops
packets when it runs out of buffer space, preventing deadlock.
However, this makes integrating RPC/TCP applications into PANIC is challenging, since it assumes
that operations occur on a packet level and the scheduler may drop an acknowledged packet, violating TCP semantics.

ClickNP~\cite{clicknp_sigcomm} is an FPGA-accelerated packet processing
framework that also supports the integration of arbitrary processing elements.
However, it does not use a NoC. Instead, components are directly connected via
FIFOs, which makes it harder to replicate elements. Since ClickNP aims to
accelerate software network functions, it lacks support for higher-level network
protocols and direct-attached accelerators. It further assumes a PCIe connection
to a CPU, which it relies on for control-plane configuration.

Rosebud~\cite{rosebud_asplos} is an FPGA framework for middleboxes. It uses an
interconnect to connect custom processing elements they call reconfigurable
processing units (RPUs) that can include accelerators. Because it targets
middleboxes, they do not evaluate a network stack with full reliable transport
protocol support. While it does provide support to chain RPUs, they acknowledge
it was not designed to do so, and inter-RPU traffic has a fairly significant
latency penalty.

A more restrictive approach leverages reconfigurable match-action tables. An
action (e.g. strip a header, rewrite a field, drop a packet) is taken based on
some header fields in the header of the packet. Typically, there is a pipeline
of these processing elements \cite{azure_smartNIC, flexnic_asplos, rmt_sigcomm}.
However, match-action style processing is not well-suited for highly stateful
processing \cite{flowblaze_nsdi} typical of application-level offloads. Other
models have been proposed for stateful packet processing.  Flowblaze uses an
FSM-based model \cite{flowblaze_nsdi}.  However, they specifically say that
workloads above the transport layer are out of scope. hXDP proposed a processor
for eBPF bytecode \cite{hxdp_osdi} designed for offloading kernel-level eBPF
programs. Because of its sequential execution model, hXDP performs best on small
programs and is a poor fit for more complex processing such as Reed-Solomon encoding. 


%

\subsection{Transport protocol offloads}
Another related vein of work are transport protocol offloads. Most of these are
TCP offload engines available as custom chips \cite{chelsio_t6} or encrypted IP
cores for FPGAs \cite{chevin_toe,mle_toe,easics_toe}. They generally do not
support the full range of functions found in datacenter network stacks.

Some TCP offload engines could potentially support modification.  Limago
\cite{limago} is an open-source TCP and RoCEv2 offload engine written in Vivado
HLS. However, it does not provide any specific APIs or hooks for adding other
protocols, so introducing a new network function or new protocol would require
fairly extensive modifications to the stack itself.  Tonic~\cite{tonic} is an
open-source implementation of the TCP send path and supports customization of
the transport protocol, but does not address any lower-level packet processing
layers; it also lacks a complete receive path implementation.  FlexTOE~\cite{flextoe_nsdi} is a software implementation of TCP offload engine using the
Netronome DPU, a processor designed specifically for network processing that is
programmable using C or eBPF. While they do support network functions, their
work targets TCP offload for CPUs while our work shows that a direct-attached
hardware accelerator does not need a CPU core to support software stack
functionality. 

Microsoft Catapult's FPGAs use a custom transport protocol called
LTL~\cite{catapult_v2}, which is a reliable transport protocol over UDP. Similar
to most TCP engines, it is presented as a fixed IP core with no interface for
extension. Catapult also supports a single-layer RMT, used for network
virtualization~\cite{azure_smartNIC}. However, it is unknown if these are ever
combined and if so, how it would support new protocols or network functions. 
\section{Conclusion} 
Modern datacenter networking relies on a variety of network functions and
protocols, but current hardware network stacks fall short on these features.  As
datacenters continue to offload computation to accelerators, it is becoming
increasingly important to enable direct-attached accelerators to reduce network
overhead. In this paper, we presented the design and implementation of
\thesystem, a NoC-based network stack for direct-attached accelerators that
is customizable and supports the variety of protocols and management
functions needed for datacenter networking. We demonstrated that \thesystem{} can
combine replicated protocol elements and replicated applications for higher
bandwidth, provide consistent low latency, with minimal overhead. We have open-sourced
\thesystem{} for reuse at \url{https://github.com/beehive-fpga/beehive}.
\section*{Acknowledgments}
We thank the anonymous reviewers for feedback that substantially improved the paper. This work was supported by grants from VMware Research, the National Science Foundation (CNS-2104548, No. 2213387), the University of Washington Center for the Future of Cloud Infrastructure (FOCI), the Intel TSA center, and the NSF Graduate Research Fellowship. Pratyush Patel assisted in an early draft of this paper.
We would like to thank Hugo Sadok for his invaluable help with getting \enso running. 
We also thank Alexey Lavrov and Jonathan Balkind for feedback on various drafts of this paper.
%
%
%
%
%

\section*{Artifact Appendix}

\subsection{Abstract}
\thesystem is a NoC-based network stack for direct attached accelerators designed to enable flexible construction of complex network functionality in hardware. 
This appendix outlines the steps to access \thesystem. All files necessary to build the various artifacts for the experiments in the paper are available in the \thesystem repository.

Full evaluation and reproduction of evaluation results requires testbed access to a 100 Gb switch, at least 4 100 Gb NICs, and an Alveo U200 FPGA. It also requires access to Vivado 2021.2 to build the hardware designs. Simulation requires access to Python and ModelSim 2019.2.


\subsection{Artifact check-list (meta-information)}

{\small
\begin{itemize}
  \item {\bf Program: }Vivado 2021.2, ModelSim 2019.2
  \item {\bf Compilation: }Follow the directions inside \verb|beehive/README.md|
  \item {\bf Hardware: }Alveo U200/U250
  \item {\bf How much disk space required (approximately)?: }200 MB
  \item {\bf Publicly available?: }Yes, see below
  \item {\bf Code licenses (if publicly available)?: }BSD 3-Clause
  \item {\bf Data licenses (if publicly available)?: }BSD 3-Clause
  \item {\bf Archived (provide DOI)?: 
  \url{https://doi.org/10.5281/zenodo.13308868}
  }
\end{itemize}
}

\subsection{Description}

\subsubsection{How to access}
\thesystem is accessible at \url{https://github.com/beehive-fpga/beehive}

\subsection{Installation}

Please refer to the instructions inside beehive/README.md

\subsection{Evaluation and expected results}

Each hardware build needs some sort of input data to run on.
You are expected to connect to the hardware via TCP or UDP and
send the appropriate packets to drive the device and to fetch
evaluation information from the device.
If simulating, the simulation driver will
drive the design. In this case, you should see outputs relating to that run.

\subsection{Methodology}

Submission, reviewing and badging methodology:

\begin{itemize}
  \item \url{https://www.acm.org/publications/policies/artifact-review-and-badging-current}
  \item \url{https://cTuning.org/ae}
\end{itemize}

\bibliographystyle{IEEEtranS}
\bibliography{tcp_hw_references,beehive_refs}

\maketitle
\end{document}